\documentclass[twocolumn, aps, longbibliography,unsortedaddress]{revtex4-2}

\usepackage{amssymb}
\usepackage[colorlinks=true,%
bookmarks=false,%
linkcolor=green,%
urlcolor=blue,%
citecolor=magenta,%
breaklinks]{hyperref}
\usepackage[english]{babel}
\renewcommand{\selectlanguage}[1]{}

\usepackage{graphicx}
\usepackage{caption}
\usepackage{booktabs}
\usepackage{subcaption}
\usepackage{dcolumn}
\usepackage{bm}
\usepackage{hyperref}
\usepackage{physics}
\usepackage{quantikz}
\usepackage{mathtools}

\renewcommand{\arraystretch}{1.5}    

\begin{document}

\preprint{APS/123-QED}

\title{Quantum Algorithm for the Fixed-Radius Neighbor Search}
\author{Luca Cappelli}
\affiliation{%
 Dipartimento di Fisica dell’Università di Trieste, via Tiepolo 11, I-34131 Trieste, Italy;\\
 Fondazione Istituto Italiano di Tecnologia, 
Center for Life Nano-Neuroscience at la Sapienza, 
Viale Regina Elena 291, 00161 Roma, Italy
\\
 INAF - Osservatorio Astronomico di Trieste, Trieste, Italy;
}%

\author{Claudio Sanavio}
\email{claudio.sanavio@iit.it}
\affiliation{Fondazione Istituto Italiano di Tecnologia, 
Center for Life Nano-Neuroscience at la Sapienza, 
Viale Regina Elena 291, 00161 Roma, Italy}

\author{Alessandro Andrea Zecchi}
\email{alessandroandrea.zecchi@polimi.it}
\affiliation{%
MOX – Department of Mathematics \\
Politecnico di Milano \\
Piazza L. da Vinci, 32, 20133 Milano, Italy
}%

\author{Philip Mocz}
\affiliation{Center for Computational Astrophysics, Flatiron Institute, 162 5th Ave, New York, NY 10010, USA}

\author{Giuseppe Murante}
\affiliation{INAF - Osservatorio Astronomico di Trieste, via Tiepolo 11, I-34131 Trieste, Italy;}

\author{Sauro Succi}
\affiliation{Fondazione Istituto Italiano di Tecnologia, 
Center for Life Nano-Neuroscience at la Sapienza, 
Viale Regina Elena 291, 00161 Roma, Italy}

\date{\today}

\begin{abstract}
Neighbor search is a computationally demanding problem, usually both time- and memory-consuming. The main problem of this kind of algorithms is the long execution time due to cache misses. In this work, we propose a quantum algorithm for the Fixed RAdius Neighbor Search problem (FRANS) based on the fixed-point version of Grover's algorithm. We propose an efficient circuit for solving the FRANS with linear query complexity with the number of particles $N$. 
The quantum circuit returns the list of all the neighbors' pairs within the fixed radius, together with their distance, avoiding the slow down given by cache miss. 

We analyzed the gate and the query complexity of the circuit. Our FRANS algorithm presents a query complexity of $\mathcal{O}(N/\sqrt{M})$, where $M$ is the number of solutions, reaching the optimal lower bound of the Grover's algorithm.
We propose different implementations of the oracle, which must be chosen depending on the precise structure of the database. Among these, we present an implementation using the Chebyshev distance with depth $\mathcal{O}(q_1)$, where $2^{q_1}$ is the number of grid points used to discretize a spatial dimension.
State-of-the-art algorithms for state preparation allow for a trade-off between depth and width of the circuit, with a volume (depth$\times$ width) of $\mathcal{O}(N\log(N))$.
This unfavorable scaling can be brought down to $\mathcal{O}(\text{poly}(\log N))$ in case of structured datasets.
We proposed a stopping criterion based on Bayes interference and tested its validity on $1D$ simulations. 
Finally, we accounted for the readout complexity and assessed the resilience of the model to the readout error, suggesting an error correction-free strategy to check the accuracy of the results.
\end{abstract}

\maketitle

\begin{figure}
    \centering
    \includegraphics[width=\columnwidth]{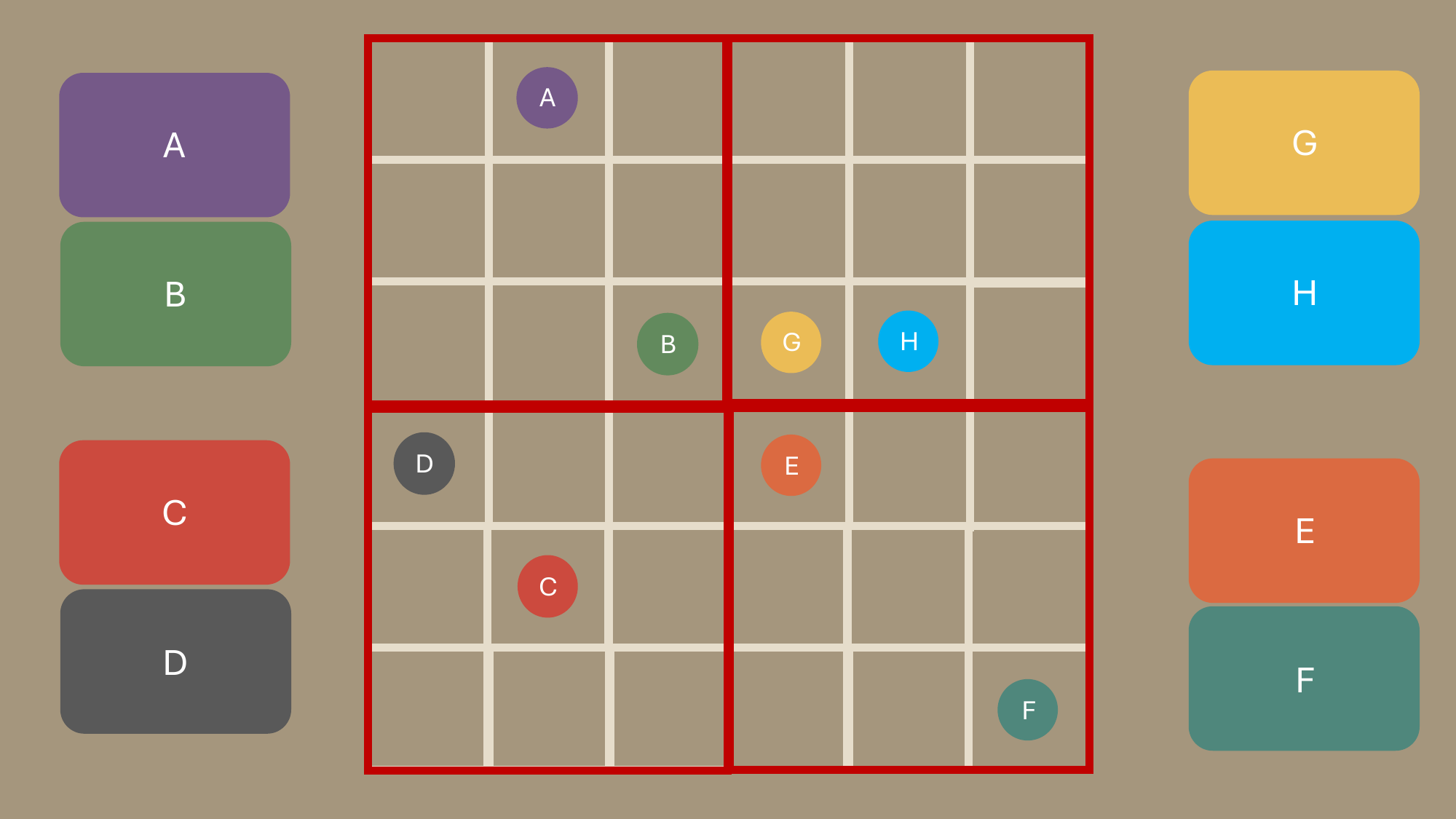}
    \caption{
    Toy model illustrating the cache miss problem in neighbor search algorithms.
    An evenly spaced grid contains eight particles distributed across four memory blocks (indicated by red boundaries).Each particle's data is stored in memory locations denoted by matching colors. 
    Note that spatially adjacent particles might be stored in different memory locations.
    }
    \label{fig: cache miss toy model}
\end{figure}

\section{Introduction}
With the ceaseless advances of processing speed, scientific computing is
entering a regime in which efficient data access policies are becoming 
prioritary with respect to sheer computing speed~\cite{KLINKENBERG2023, DRAVAI2025}. 
Accessing data can be more costly than processing them, especially in applications based on irregularly distributed or unstructured data, such as particle methods and Lagrangian formulations of continuum theory, where physical and logical (storage location in memory) distances often diverge. Unstructured data also appear in Eulerian fields in disordered environments, such as flows in porous media~\cite{feder2022, sahimi1993, cali1992}.

A paradigmatic example of this challenge is the Fixed-Radius Neighbor Search (FRANS)~\cite{turau_fixed-radius_1991}, which consists in finding all particles within a fixed distance between each other. This Fixed-RAdius Neighbor Search (FRANS)~\cite{turau_fixed-radius_1991} is strictly related to $N$-body methods and forms the computational backbone of numerous simulation methods across physics, chemistry, and astronomy~\cite{verlet_computer_1967, SPH_OG, Gadget, gromacs, DiIlio_fluiddynamics_2018}, yet consistently dominates runtime despite decades of algorithmic advancements.

The naive approach to finding all particles within a cutoff distance $\xi$ requires examining every possible particle pair, resulting in $\mathcal{O}(N^2)$ scaling that quickly becomes unfavorable as the system size grows. Modern simulations employ specialized spatial data structures such as uniform grids, Verlet lists~\cite{verlet_computer_1967, Verlet_list1, gromacs}, and hierarchical trees~\cite{burnes-hut-tree, Gadget}, which can reduce the complexity to $\mathcal{O}(N \, \log N)$ by  
considering only a portion of all the possible pairs. 

Fast algorithms used to explicitly find all the particle pairs within the fixed-radius scale as $\mathcal{O}(M+N)$, where $M$ is the number of neighboring pairs, with a preprocessing of $\mathcal{O}(N\log N$) and a $\mathcal{O}(N)$ space complexity~\cite{chen_fast_2024}.
Yet, in spite of the improved computational cost, this optimized neighbor search still constitutes the primary computational expense across diverse scientific applications. 

In the context of Molecular Dynamics (MD), Verlet lists or cell indices~\cite{verlet_computer_1967} are used to evaluate interactions between atoms for the study of a wide range of phenomena such as drug binding~\cite{drug_binding1, drug_binding2}, membrane dynamics~\cite{membrane_dynamics1, membrane_dynamic_2} and material science~\cite{polymer_dynamics}. 
These data structure keep track of the atoms that are spatially close, in order to avoid considering atoms far beyond the interaction radius. The FRANS procedure and maintenance of these lists typically consumes from $30\%$ to $60\%$ of total runtime in large-scale simulations~\cite{Verlet_list_time, verlet_gpu_comparison_2}. 
Another interesting field of MD concerns protein folding. Research in this area focuses on how proteins evolves towards an equilibrium state (folding) and interact with one another. A noticeable application is found in neuroscience, where abnormal folding patterns often signal neurological diseases~\cite{protein_folding_and_neuro_diseases}. While earlier studies relied on neighbor search~\cite{LindorffLarsen2011}, the field has evolved considerably with new approaches. AlphaFold~\cite{alphafold3} stands out as a particularly valuable tool, serving as a detailed simulator and providing a database of $200$ million proteins' structure. In this case, neighbor search algorithms remain useful for homology detection~\cite{homology1}, helping researchers identify structural similarities between proteins to trace evolutionary origins and primitive forms. These methods also assess how well different proteins might interact based on their geometric properties~\cite{homology2}.
Non-equilibrium protein folding~\cite{non_eq_pf_2, non_eq_pf_3, Nonequilibrium1pf} is an open research field, also related to neurological disorder~\cite{non-eq-for-neuro1, non-eq-for-neuro2} and represents another area of interest for neighbor search. Here environmental influences cause proteins to adopt conformations outside thermal equilibrium. Since ground-state assumptions do not apply in these systems, different methods are used to model the folding mechanism such as coarse-grained molecular dynamics simulations combined with FRANS.

In both cosmology and fluid dynamics, Smoothed Particle Hydrodynamics (SPH)~\cite{SPH_OG} can be used to simulate the evolution of fluids and gases. This approach represents continuous media as discrete particles, each carrying fundamental properties such as mass, momentum and energy. Particle interactions emerge through a  weighting scheme that employs localized kernel functions to determine the influence of neighboring particles. The method achieves conservation of field quantities (e.g, pressure and density) by spatially averaging contributions from all particles within the kernel's smoothing length radius.
The evaluation of neighboring pair and interaction between particles takes up most of the execution time~\cite{sph_today_time}. The challenge becomes particularly acute in simulations with free surfaces or multiphase flows, where particle distributions become highly non-uniform. 
\\
In the context of astrophysical N-Body simulations, SPH is used to simulate dust dynamics using tree-based methods for the neighbor search part.
It is common to consider in these scenarios the nearest $n$ neighbors (typically $n=64$) rather than all particles within a fixed radius~\cite{springel2005cosmological}. While this can introduce limitations to the convergence of the scheme~\cite{zhu2015numerical}, it remains a practical approach. 
\\
For purely gravitational N-body simulations, hybrid Tree-Particle-Mesh methods are often employed to achieve $\mathcal{O}(N \log N)$ scaling, leveraging techniques such as space-filling Peano-Hilbert ordering, Oct-trees, and multi-pole expansions. However, these methods do not explicitly compute interactions for all neighbor pairs within a fixed radius, and thus do not fully fall into the FRANS category.
In both cases, the traversal of these spatial data structures to identify nearby particles still dominates computational expense, consuming most of simulation's time in cosmological models with billions of bodies~\cite{tree_traversal_time, Gadget}. Highly clustered mass distributions -- characteristic of galaxy formation -- further amplify these costs by necessitating frequent tree rebuilds.

What makes this problem particularly resistant to optimization is not merely its algorithmic complexity, but rather the challenging computational patterns it creates. Specifically, the combination of a large data structure combined with the frequent and irregular memory access patterns. 

To be more specific, modern computing architectures are built around a memory hierarchy (CPU registers $\rightarrow$ L1 cache ($32-64 \, \text{KiB}$ per core) $\rightarrow$ L2 cache ( $512 \text{KiB}$ $-8\,\text{MiB}$ per core) $\rightarrow$ L3 cache ($120-256\,\text{MiB}$ per core) $\rightarrow$ main memory; data collected from~\cite{superpc1, superpc2}) that performs best when data access follows predictable patterns. 
However, neighbor search fundamentally involves unpredictable memory access and particles that are close in space configuration might be stored far apart in memory, particularly after many simulation time steps have caused the particles to move from their initial positions. This creates numerous cache misses, as the processor constantly needs to fetch data from slower memory tiers. 
For reference, in both SPH and MD the number of   $M$ depends on the application, as high resolution simulations might require higher values. In the common scenario however,  an accurate $3$D SPH simulation requires approximately $N = 10^{10}$, with approximately $400$ target neighbors per particle~\cite{SPH_neighbors},  while an MD simulation might require $N=10^8$ with the number of neighbors in the order of hundreds~\cite{castelli_decrypting_2024,Frenkel2002, gromacs}.

In Fig.~\ref{fig: cache miss toy model} is presented a toy model representing the cache miss problem.
When searching for neighbors of particle B, the algorithm needs to access data for particles E, G, and H that happen to be spatially nearby. However, due to limited cache memory capacity, only particles within B's sub-quadrant are currently stored in cache. To retrieve information for the remaining particles, the algorithm must fetch data from main memory—a significantly slower operation. The frequent repetition of this process creates substantial computational overhead and dramatically increases execution time.
Thus, the mismatch between physical proximity and memory proximity causes the FRANS to be the bottleneck in many different scenarios.
We note that the toy model in Fig.~\ref{fig: cache miss toy model} represents the cache-miss problem in its simplest form. In practice, sophisticated techniques exist to reduce the mismatch between physical and memory proximity, such as ordering particles along Peano-Hilbert space-filling curves~\cite{alves2022cache, springel2005cosmological}.

An alternative approach to solving the bottleneck problem can be offered by the framework of quantum computation~\cite{quantum_for_many_body} which in the last two decades has seen a surge of theoretical advances and potential applications~\cite{nielsen_quantum_2010} in quantum chemistry~\cite{cao_quantum_chemistry_2019}, material science~\cite{alexeev_quantum-centric_2024}, cosmology~\cite{cappelli_vlasov_2024} and computational fluid dynamics~\cite{Succi_2023_overview}.

At the dawn of quantum computing, Grover's search algorithm~\cite{grover_fast_1996} showed one of the first example of a quantum algorithm that can run faster than any classical algorithm designed for the same task. 
While the computational complexity of a classical search is $\mathcal{O}(N)$, Grover's algorithm can do the search with only $\mathcal{O}(\sqrt{N})$ steps, thus offering a quadratic advantage. 
After its appearance in the literature, Grover's algorithm has been analyzed and modified several times to match more specific tasks and to tackle the inherited issues of the original algorithm. 
Among those, we mention the Fixed-point search (FPS) algorithms~\cite{grover_fixed-point_2005,mizel_critically_2009,yoder_fixed-point_2014}, thought to tackle the periodic increase and decrease in the success probability of the search, and the oblivious amplitude amplification algorithms(OAA)~\cite{berry_exponential_2014,yan_fixed-point_2022,zecchi_improved_2025} intended to overcome the issue of not knowing the target state in advance. 
We will discuss both approaches later in the main text. 
In more recent years, it has been shown how Grover's search algorithm and all of its many versions belong to the class of Quantum Singular-value Transform (QSVT) algorithms~\cite{gilyen_quantum_2019,martyn_grand_2021}, which reveals the principles behind the most widely used (and cited) quantum algorithms.

In this work, we aim to find a solution to the FRANS problem by employing our own version of the oblivious-fixed-point amplification algorithm~\cite{mizel_critically_2009,yan_fixed-point_2022}, which is very adaptable to different initial states while being agnostic to the target state.
The manuscript is structured as follows. In Section~\ref{sec:II} we introduce the problem and describe its difficulties when treated by a classical computer. In Section~\ref{sec:III} we describe the Grover's algorithm and its FPS and OAA versions. In Section~\ref{sec:IV} we describe our quantum FRANS algorithm and we highlight the similarities and differences with the previous algorithms. Hence, we explicitly show the quantum circuit and the gate decomposition of the various operations, analyzing the computational complexity and the gate complexity. Furthermore, we propose a stopping criterion based on Bayesian's posterior probability.
In Section~\ref{subsec: numerical tests} we perform numerical test to verify the goodness of our stopping criterion.
In Section~\ref{sec:V} we introduce the bit-flip and the readout measurement errors in our simulation, calculating the threshold for the amount of noise required to preserve the quantum advantage of the algorithm, without introducing particular error correction schemes. Finally, we draw the outlooks and the conclusions of our work in Section~\ref{sec:VI}.

\section{The fixed-radius neighbor search}\label{sec:II}

At the heart of many computational models in physics lies a conceptually simple yet computationally intensive operation: determining which objects in space are close to one another. This procedure, known as neighbor search, forms the foundation for virtually all particle-based simulation methods across the physical sciences.

In its most fundamental form, the fixed-radius variant, neighbor search answers a straightforward question: given a collection of particles distributed throughout space, which particles lie within a fixed distance of each other? This critical calculation enables the modeling of interactions that occur only between objects in close proximity—forces that diminish rapidly with distance, collisions between bodies, or influence that spreads within a limited radius.

The mathematical formulation is deceptively simple. For each particle $i$ in a system of $N$ particles, we must identify all other particles $j$ such that the distance between them $d(r_i, r_j)$ is less than some threshold distance $\xi$. 
This distance threshold might represent the cutoff radius of a potential energy function in MD, the smoothing length in fluid simulations.

Despite its conceptual simplicity, this operation presents an extraordinary computational challenge. The most direct approach—checking every possible pair of particles—would require examining $N^2$ distance calculations, a prohibitive scaling for systems of scientific interest that may contain millions or billions of particles. Moreover, as particles move throughout a simulation, these neighborhood relationships continually change, requiring repeated recalculation.

To address this challenge, researchers have developed sophisticated spatial data structures that organize particles based on their positions in space. Cell lists divide the simulation domain into a grid, allowing the search to focus only on particles in adjacent cells. Verlet lists maintain a precomputed set of neighbors for each particle, including a buffer zone to reduce update frequency. Tree-based methods hierarchically partition space, enabling rapid elimination of distant regions from consideration.

These techniques reduce the theoretical complexity from $O(N^2)$ to $O(N \log N)$, representing an enormous computational saving. Nevertheless, neighbor search still typically accounts for $30-70\%$ of total computation time in production simulations across disciplines~\cite{verlet_gpu_comparison_2, sph_today_time, Verlet_list_time}. The persistent challenge stems from a fundamental disconnect between two different types of proximity: spatial proximity in the physical simulation and memory proximity in the computing hardware.

When particles are close to each other in the simulation space, their data should ideally be close to each other in the computer's memory for efficient processing. However, this alignment rarely occurs, especially as particles move throughout the simulation. A particle's neighbors in physical space are often scattered across distant memory locations, forcing the processor to constantly fetch data from widely separated memory addresses.

Compounding this problem is the repetitive nature of the search operation. For each of the $N$ particles in the system, the algorithm must perform a separate neighbor search, repeatedly accessing memory locations across the entire data structure. 

The combined effect of these two factors -- spatial-memory misalignment and repeated broad memory access -- frequently results in cache misses, where the processor must wait to retrieve data from slower memory tiers rather than finding it in fast cache memory. These waiting periods significantly increase actual execution time, often by an order of magnitude or more compared to theoretical predictions.

Thus, the primary challenge in modern neighbor search implementations has shifted from algorithmic complexity to hardware efficiency. Even algorithms with optimal theoretical scaling ($O(N)$) can perform poorly in practice due to their incompatibility with contemporary computing architectures and the true complexity now lies not in the mathematical operation count but in navigating the complex memory hierarchy of modern processors to minimize data movement costs.

Estimating the impact of cache misses on the FRANS solution is a complex issue that falls beyond the scope of this work. Such an assessment depends not only on the algorithmic implementation but also on the underlying hardware architecture and specific software realization. Additionally, factors such as dataset size, memory management strategies, and data distribution across computing resources vary significantly between different versions and implementations. Nevertheless, to provide context on how this issue impacts modern simulations, we present a simplified estimate.
\\
Consider a cosmological simulation with $10^{10}$ particles executed on Fugaku supercomputer~\cite{superpc1}. It has been reported that in a $32^3$ particles SPH  simulation optimized for single-node execution on Fugaku exhibits cache miss probabilities for $L1 = 1.2\%$ and for $L2 = 0.12\%$~\cite{eurex}. Building on these data, we aim to estimate cache behavior for a $10^{10}$-particle simulation under a simplified capacity-dominated scaling model, where the miss rate scales proportionally with the number of particles per node.
\\
Fugaku comprises 158,976 compute nodes~\cite{superpc1}. A reasonable allocation for a simulation of this magnitude would utilize approximately one-tenth of the available nodes ($\sim$15,898 nodes)~\cite{Habib2017HACC}, with each node processing $\sim 6.29 \times 10^{5}$ particles. Applying a naive linear scaling law, the estimated miss rates increase to $L1 \simeq 23\%$ and $L2 \simeq 2.4\%$.
\\
These estimates provide a first-order approximation of cache behavior, assuming that the per-particle data dimension remains independent of the working set size. However, for larger datasets, a significantly greater fraction of particles resides in RAM compared to the $32^3$ particles-per-node case, thereby increasing the probability of accessing particles stored in main memory rather than cache. This type of miss is particularly costly, as retrieving data from main memory requires substantially more CPU cycles than cache access. During such memory fetches, the entire computational pipeline stalls while awaiting the memory refill operation.
\\

Given these considerations, we regard our estimate as an upper bound on the actual cache miss percentage. We emphasize that actual performance may differ substantially depending on code-specific factors, particularly the strategies employed for managing and accessing larger datasets in practice. For instance, codes written for molecular dynamics do not translate well to the astrophysical domain due to fundamentally different data layouts: Verlet lists are optimized for short-range interactions, whereas in cosmological simulations data are clustered in distinct spatial regions separated by cosmic voids. Moreover, state-of-the-art codes employ sophisticated techniques such as Peano-Hilbert space-filling curves~\cite{alves2022cache, grover_fixed-point_2005} to ensure that spatially proximate data occupy adjacent memory locations, thereby minimizing cache misses. However, the focus of this work is not to compete with these highly optimized classical methods. Rather, we aim to introduce the fundamental elements of a quantum algorithm for the FRANS problem.

\section{Quantum search algorithms}\label{sec:III}

In this section, we describe a few versions of the Grover's algorithm as they are relevant to this paper. 
In the original work, the $N$ elements of a database $D$ are encoded into a quantum state $|i\rangle$ and we want to select a specific target state $|\psi_T\rangle$. 
The initial state is set as the uniform superposition of all possible states 

\begin{equation}\label{eq:uniform_superposition}
    |\psi_0\rangle = \frac{1}{\sqrt{N}}\sum_{i\in D}|i\rangle.
\end{equation}

The algorithm consists in applying to the state $|\psi_0\rangle$ first an oracle operator $\hat{O}$, which is capable of recognizing the solution and whose action is to flip the sign of the target state $|\psi_T\rangle$,  defined as

\begin{equation}\label{eq:oracle_operator}
\hat{O} = \mathbb{I}-2|\psi_T\rangle\langle\psi_T|,
\end{equation}

\noindent and later applying the reflection operator $\hat R_0$, which reflects the resulting state with respect to the initial state $|\psi_0\rangle$, namely

\begin{equation}\label{eq:reflection_operator}
\hat{R}_0 = 2|\psi_0\rangle\langle\psi_0|-\mathbb{I}.
\end{equation}

Conveniently, the algorithm can be represented in a two-dimensional space spanned by the solution $|\psi_T\rangle$ and its orthogonal part $|\psi^\perp\rangle$ in the initial state, defined such that
\begin{equation}
    |\psi_0\rangle = \cos\theta|\psi^\perp\rangle+\sin\theta|\psi_T\rangle,
\end{equation}

\noindent where the angle $\theta=\arcsin\bigg{[}\sqrt{M/N}\bigg{]}$ and $M$ is the number of target states in the database. 
The application of the Grover operator $\hat{G}=\hat R_0\hat O$ rotates the vector in the $|\psi_T\rangle,|\psi^\perp\rangle$ plane by an angle $2\theta$ increasing the amplitude of the target state. 
By applying $G$ a sufficiently large number of times $k\sim\sqrt{N/M}$, the final state $G^k|\psi_0\rangle$ will have the maximum amplitude for the component $|\psi_T\rangle$. 

Now we see that there are two main hurdles when one aims to apply the Grover's search algorithm to its database. 
First, if the value $M$ is unknown, the risk is to apply $G$ too many times, with the result that the final state passes over the $|\psi_T\rangle$ axis and the amplitude probability of measuring the target state reduces.
This is known as the \textit{souffl\'e problem}, since applying the Grover's operator too few times undercooks the state, whereas applying it too many times overcooks it, with the resulting deflation of the quantum cake. This obstacle was overcome by the FPS algorithm~\cite{mizel_critically_2009}, which increases monotonically the probability of measuring a target state.
Second, as expressed in Eq.~\eqref{eq:oracle_operator}, the target state has to be known in order to construct the correct oracle operator $\hat{O}$. Although this may be sensible in the context of quantum simulation to construct a particular quantum state, it loses its significance when the quantum algorithm is intended to search over a classical database or to solve classical optimization problems. The OAA algorithm was designed to overcome this problem. 
In the remainder of this section we are going to summarize these two algorithms.

\subsection{FPS algorithm}

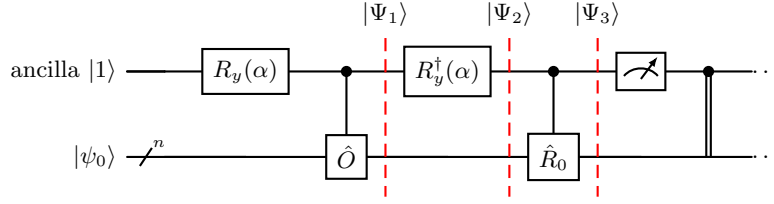
\begin{figure}
    \centering
    \resizebox{\columnwidth}{!}{%
    \begin{quantikz}
        \lstick{ancilla $\ket{1}$} & \qw & \gate{R_y(\alpha)} & \ctrl{1}\slice{$|\Psi_1\rangle$} & \gate{R_y^\dagger(\alpha)}\slice{$|\Psi_2\rangle$} & \ctrl{1}\slice{$|\Psi_3\rangle$} & \meter{} & \ctrl{0} & \dots\\
        \lstick{$\ket{\psi_0}$} & \qwbundle{n} & \qw & \gate{\hat{O}} & \qw & \gate{\hat R_0} & \qw & \wire[u][1]{c} & \dots
    \end{quantikz}%
    }
    \caption{Circuit used for the amplitude amplification process. Based on Mizel search algorithm. The parameter $\alpha$ is adjusted accordingly to the iteration. The measurement on the ancilla qubit serves as control: if $\ket{0}$ is the outcome, it means that the quantum register is ready for measurement; otherwise repeat the process, varying the angle $\alpha$.}
    \label{fig:Mizel_algorithm}
\end{figure}

The FPS algorithm~\cite{mizel_critically_2009} introduces an ancilla qubit to reproduce a non-unitary dynamics that damps out the oscillations of the results of Grover's algorithm between the target and non-target states. The algorithm uses a series of parametric rotations
\begin{equation}
    R_y(\alpha_i) = \begin{pmatrix}
        \cos\frac{\alpha_i}{2} & -\sin\frac{\alpha_i}{2}\\
        \sin\frac{\alpha_i}{2} & \cos\frac{\alpha_i}{2}
    \end{pmatrix},
\end{equation}

\noindent where the optimal value of the $\alpha_i$ angles depends on the specifics of the problem. 
The algorithm is presented in Fig.~\ref{fig:Mizel_algorithm} and reads as follows

\begin{align}\label{eq:FPSA}
&\Pi_{i=1}^{\text{K}}\big{[}\big{(}|0\rangle\langle0|\otimes \mathbb{I}+|1\rangle\langle1|\otimes \hat{R}_0\big{)}\big{(}R_y(-\alpha_i)\otimes \mathbb{I})\big{)}
\\&\nonumber
\big{(}|0\rangle\langle0|\otimes \mathbb{I}+|1\rangle\langle1|\otimes \hat O\big{)}\big{(}R_y(\alpha_i)\otimes \mathbb{I})\big{)}\big{]}|1\rangle|\psi_0\rangle,
\end{align}

\noindent where the circuit is run a total number of $\text{K}$ times depending on the result of a measurement performed on the ancilla qubit as we see next.

The first step of the algorithm rotates the ancilla qubit to $-\sin\frac{\alpha_i}{2}|0\rangle+\cos\frac{\alpha_i}{2}|1\rangle.$ The second term performs a controlled operation which yields the state 

\begin{equation}
|\Psi_1\rangle = -\sin\frac{\alpha_i}{2}|0\rangle|\psi_0\rangle+\cos\frac{\alpha_i}{2}|1\rangle(\cos\theta|\psi^\perp\rangle-\sin\theta|\psi_T\rangle).
\end{equation}

The application of the second single-qubit rotation $R_y(-\alpha_i)$ leads to

\begin{align}
|\Psi_2\rangle &= -\sin\alpha_i\sin\theta|0\rangle|\psi_T\rangle +\sin^2\frac{\alpha}{2}|1\rangle|\psi_0\rangle
\\ \nonumber
&+\cos^2\frac{\alpha}{2}|1\rangle\big{(}\cos\theta|\psi^\perp\rangle-\sin\theta|\psi_T\big{)} 
\\ & \nonumber
= -\sin\alpha_i\sin\theta|0\rangle|\psi_T\rangle-\cos\alpha_i\sin\theta|1\rangle|\psi_T\rangle
\\ & \nonumber
+\cos\theta|1\rangle|\psi^\perp\rangle.
\end{align}

Now the action of the last controlled operation ${\big{(}|0\rangle\langle0|\otimes \mathbb{I}+|1\rangle\langle1|\otimes \hat{R}_0\big{)}}$ on $|\Psi_2\rangle$ is given by

\begin{align}\label{eq:FPS_results}
|\Psi_3\rangle= & -\sqrt{p_{i+1}}|0\rangle(|\psi_T\rangle)
\\ & \nonumber
+\sqrt{1-p_{i+1}}|1\rangle\big{(}(c_i\cos2\theta- s_i\cos\alpha_i\sin2\theta)|\psi^\perp\rangle
\\ & \nonumber
+(c_i\sin2\theta + s_i\cos\alpha_i\cos2\theta)|\psi_T\rangle\big{)},
\end{align}

\noindent where we have used $p_{i+1}=\sin^2\alpha_i\sin^2\theta$, $c_i=\cos\theta$ and $s_i=\sin\theta$.
Finally, depending on the outcome of the measurement performed on the ancilla qubit, we decide to stop or continue the algorithm. In fact, if we measure the state $|0\rangle$ on the ancilla qubit, the system collapses into the state $|\psi_T\rangle.$ 
Conversely, if the outcome is $|1\rangle,$ the system becomes a new superposition of the states $|\psi_T\rangle,|\psi^\perp\rangle$. 

At the $i$-th repetition, the superposition state associated with $|1\rangle$ is

\begin{equation}\label{eq:change_state}
   |\psi_{i+1}\rangle = c_{i+1}|\psi^\perp\rangle +s_{i+1}|\psi_T\rangle,
\end{equation}

\noindent with the new coefficients defined as

\begin{eqnarray}\label{eq:sincos_FPS}
    s_{i+1} &=&\frac{c_i\sin2\theta + s_i\cos\alpha_i\cos2\theta 
    }{\sqrt{1-p_{i+1}}}\\
    c_{i+1} &=&\frac{c_i\cos2\theta- s_i\cos\alpha_i\sin2\theta
    }{\sqrt{1-p_{i+1}}}\nonumber
\end{eqnarray}

\noindent for $i=1,\dots,\text{K}$ and 
\begin{equation}\label{eq:instantaneous_prob}
    p_{i+1} = s_i^2\sin^2\alpha_i,
\end{equation} 
being the probability of measuring the ancilla qubit on $|0\rangle$.

Note that, depending on the choice for the sequence of the angles $\alpha_i$, the instantaneous probability $p_i$ may oscillates or even reduce at each iteration, but the cumulative probability $p_{\text{cum}}$, that is the probability of getting the outcome $|0\rangle$ at least once after after $i$ iterations, is always non-decreasing. 
This is obtained by considering the probability of the complementary event, i.e. not measuring the desired outcome in the first $i$ iterations.

\begin{equation}\label{eq:cumulative}
    p_{\text{cum}}(i)=1-\prod_{j=1}^i\big(1-p_{j+1} \big)\quad\text{for } i=1,...,K
\end{equation}

\noindent The average number of oracle calls before success is given by:
\begin{equation}
    <n_{\text{calls}}>=\sum_i i\times \big[p_{i+1}\prod^{i-1}_{j=1}(1-p_{j+1}) \big],
\end{equation}
\noindent where the term between squared brackets is the probability of having success exactly at the $i$-th iteration.

A sensible choice of the $\alpha_i$ leads the cumulative probability of success to rapidly grow to 1. In the case of knowing beforehand the value of $M$, and consequently of $\theta$, the best choice is given by the constant critical value \begin{equation}
    \alpha_C=\arccos\big(\frac{1-\sin2\theta}{1+\sin2\theta}\big),
\end{equation} which has been found to be the optimal angle dampening the oscillations of Grover's algorithm between target and nontarget states\cite{mizel_critically_2009}. When $M$ is not known in advance but $M\ll N$, then the critical angle computed for $M=1$ is an effective choice for a large set of values of $M$ near 1. It is noteworthy to mention that the quantum counting algorithm has been proposed as a method to efficiently count the number of solutions that attain specific requirements of a given quantum search problem or simply counting the total number of solutions \cite{brassard1998quantum}. More recently, a new method has been proposed to estimate $M$ by exploiting the relationship between this value and the probability of measuring one solution after a given number of iterations of the standard Grover's algorithm \cite{electronics13234830}.

\begin{figure*}[t] 
    \centering
    \resizebox{1.05\linewidth}{!}{
    \begin{subfigure}[b]{0.45\linewidth} 
        \includegraphics[width=\linewidth]{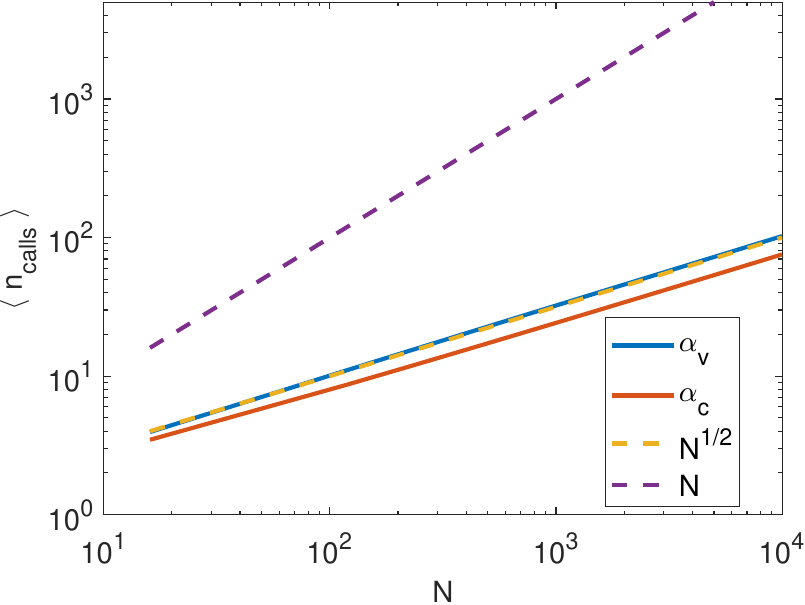}
        \caption{} 
        \label{fig:scaling} 
    \end{subfigure}
    \hfill 
    \begin{subfigure}[b]{0.45\linewidth}
        \includegraphics[width=\linewidth]{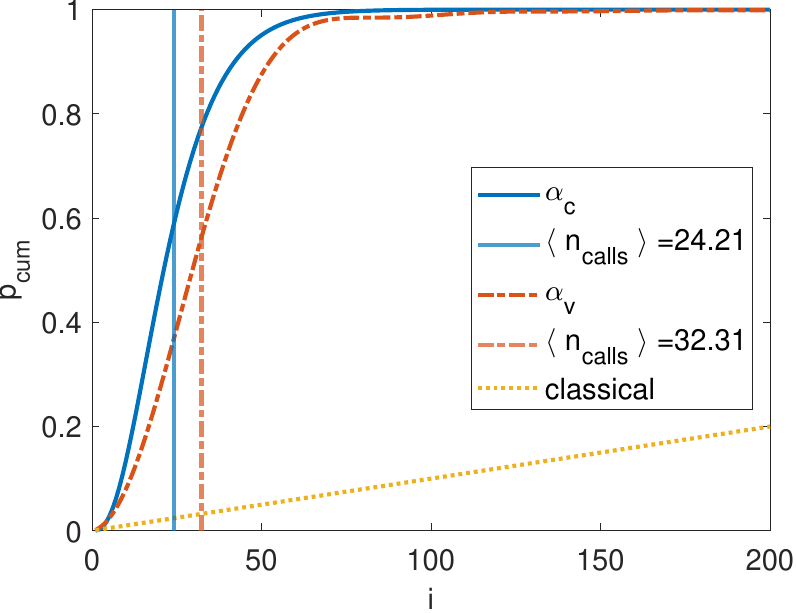}
        \caption{}
        \label{fig:comparison}
    \end{subfigure}


    \begin{subfigure}[b]{0.45\linewidth}
        \includegraphics[width=\linewidth]{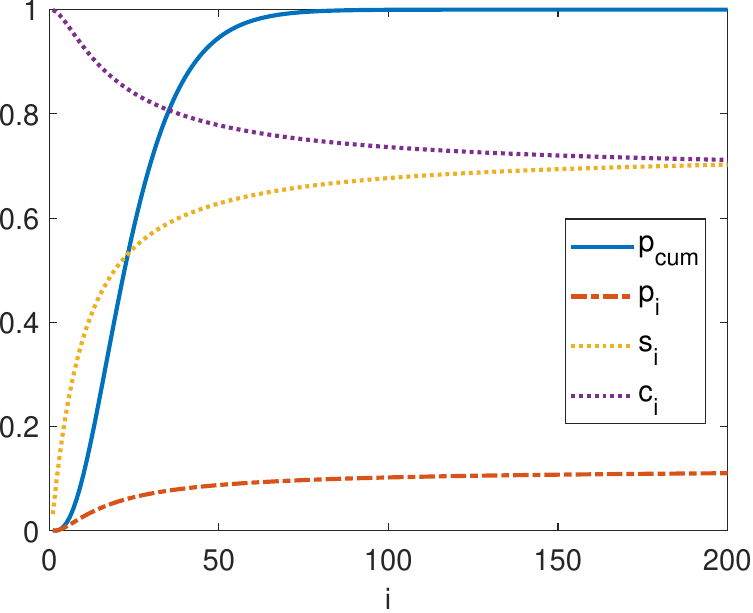}
        \caption{}
        \label{fig:critical}
    \end{subfigure}
    \hfill
    \begin{subfigure}[b]{0.45\linewidth}
        \includegraphics[width=\linewidth]{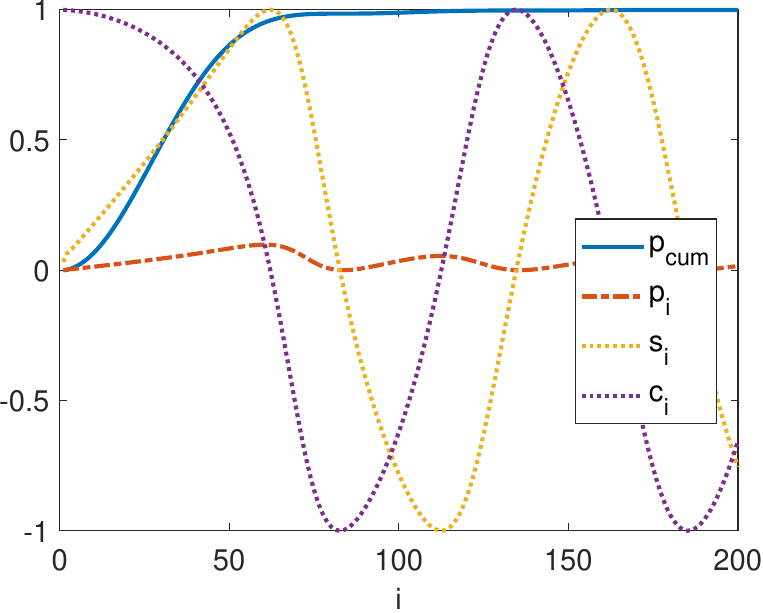}
        \caption{}
        \label{fig:descending}
    \end{subfigure}
    }
    \caption{In (a), the average number of oracle calls before success (using $M=1$ and changing the database size $N$) for the constant critical value of $\alpha_C=\alpha_i$ and for the variable decreasing sequence, defined in \eqref{eq:varying_alpha} and referenced as $\alpha_V$. This is compared with the references $\sqrt{N}$ and $N$ using a logarithmic scale on both the horizontal and vertical axes. The fixed point search algorithm provides a quadratic advantage with respect to classical search, for the considered angle's schedules. In (b), a comparison between the cumulative probabilities for the two choices with respect to a classical algorithm. The vertical lines represent the average number of queries to the oracle for the respective cases. In (c) and (d) the time evolution of the coefficients in Eq.~\eqref{eq:sincos_FPS}, the instantaneous and cumulative probabilities $p_i, p_{\text{cum}}$ for the case $M=1$, $N=1000$ with $\alpha_C$ and $\alpha_V$ respectively.  }
\end{figure*}

We show next, for the case of $M=1$ and $N=1000$, the evolution of the coefficients $c_i,s_i$ together with the instantaneous and cumulative probabilities $p_i,p_\text{cum}(i)$ for different choices of $\alpha_i$. In the case of using a constant critical angle, shown in Fig.~\ref{fig:critical}, the probability of success at each iteration $p_i$ tends to increase. 
However, when the number of target states $M$ is not known but it is supposedly very large, an effective choice is given by varying the angle in the following decreasing way 
\begin{align}\label{eq:varying_alpha}
    &\alpha_1=\pi/2, 
    \\ \label{eq:varying_alpha} 
    &\alpha_i=\ \arccos \big(\frac{1-\sin(\pi/2i)}{1+\sin(\pi/2i)}\big) \quad\text{for } i>1,
\end{align} 

\noindent for which the dynamics is plotted in Fig.~\ref{fig:descending}. In Fig.~\ref{fig:comparison}, we compare the cumulative probabilities  and the average number of oracle calls $<n_{calls}>$ before success (vertical lines) for different choices of $\alpha$ (critical and decreasing) with the probability of classical random extraction (dashed line). Comparing the two different choices of $\alpha$, the cost of not knowing $M$ weights at most by a factor of $1.5$ in the average number of oracle calls. Summing everything up, in Fig.~\ref{fig:scaling} we compare the average number of oracle calls for the case $M=1$ as a function of $N$ for the critical and decreasing cases with respect to the references $\sqrt{N/M}$ and $N/M$. As expected, we find that the fixed point search algorithm has a scaling $\mathcal{O}(\sqrt{N/M})$, an improvement with respect to the classical $\mathcal{O}(N/M)$

\subsection{OAA algorithm}

In order to apply either the Grover or the FPS algorithm, we need to know how to prepare the oracle $\hat{O}$ as expressed in Eq.~\eqref{eq:oracle_operator}, which assumes some knowledge of the target state $|\psi_T\rangle$. Most of the times, we do not have all the pieces of information about the target state, but some of its values, which are stored in a subspace that labels if the quantum state is (or it is not) the target. Namely, we can think of a unitary operation $\hat U$ such that

\begin{eqnarray}
        \hat U|0\rangle|\psi\rangle &=& |0\rangle V|\psi\rangle+|\Phi^\perp\rangle
\end{eqnarray}

\noindent where $|\psi_T\rangle=V|\psi\rangle$ has the label qubit set to $|0\rangle$ and $|\Phi^\perp\rangle$ is a state orthogonal to $|0\rangle|\psi_T\rangle$ with different value of the label qubit.
The OAA algorithm allows the amplification of the state labeled by $|0\rangle$ regardless of the quantum state $|\psi_T\rangle.$ This is a typical condition that we encounter in block-encoding algorithms~\cite{gilyen_quantum_2019} where the relevant dynamics is labeled by the $|0\rangle$ state.

We refer to the original paper~\cite{berry_exponential_2014} for the description of the algorithm. However, as a caveat, we remind that the OAA algorithm is exact only if the $V$ operator is unitary~\cite{zecchi_improved_2025}, otherwise introducing errors into the target state.

\section{Quantum FRANS algorithm}\label{sec:IV}
In this section we present our Quantum algorithm for the FRANS problem (QFRANS), which is obtained as  an efficient and ad-hoc modification of the FPS algorithm. Later in this section we present a full analysis of each component of the quantum circuit.

We consider a dataset $\mathbb{X}=\{i,x_i\}_{i=1}^N$ which collects the label $i$ given to $N$ particles and the respective position $x_i$. Our goal is to find all the pairs $(i,j)$ such that the distance $d(i,j)$ is lower than the chosen fixed radius $h$.

The first part of the algorithm encodes two copies of the dataset in four quantum registers, by using bit-encoding. In particular, two quantum registers with $q_0 = \lceil\log_2N\rceil$ qubits are used to encode the label of the particles, with the state $|i\rangle$ being the binary representation of the integer $i$, while $q_1$ qubits are used to encode their position by discretizing the whole space into a $2^{q_1}$-points lattice, and letting the state $|x_i\rangle$ to represent the coordinates. 

The preparation of the state is accomplished by employing the $\text{PREP}$ operator, which acts on the position and label registers as

\begin{equation}
    \label{eq:PREP_operator}
    \ket{\phi} = \text{PREP}|0\rangle_{q_0}|0\rangle_{q_1}=\frac{1} {\sqrt{N}}\sum_i\ket{i}_{q_0}\ket{x_i}_{q_1},
\end{equation}
\noindent where the suffix given to the ket details the number of qubits. QFRANS applies $\text{PREP}$ to both the particles register, yielding

\begin{align}
    \label{eq: initial partial state}
    \ket{\psi_0} &=\text{PREP}\otimes\text{PREP}|0\rangle_{q_0}|0\rangle_{q_1}|0\rangle_{q_0}|0\rangle_{q_1} 
    \\ & \nonumber
    = \frac{1}{N}\sum_{i,j} \ket{i}_{q_0}\ket{x_i}_{q_1}\ket{j}_{q_0}\ket{x_j}_{q_1},
\end{align}

Once the dataset is encoded we evaluate the distance between all the points in the dataset $d_{ij}=d(x_i, x_j)$. This can be done by introducing a set $a_1$ of ancillary qubits initialized in $|0\rangle$ and by applying a 
generic distance operator $\hat{D}$ on the two $q_1$-qubits registers

\begin{align}
    \label{eq: initial state}
    \ket{\psi_1} &=(\mathbb{I}_{2q_0}\otimes\hat{D})|\psi_0\rangle|0\rangle_{a_1} 
    \\ & \nonumber
    = 
    \frac{1}{N}\sum_{i,j}\ket{i}_{q_0}\ket{x_i}_{q_1}\ket{j}_{q_0}\ket{d_{ij}}_{q_1}|\cdot\rangle_{a_1};      
\end{align}

an efficient implementation of the operator $\hat{D}$ in case of euclidean geometry is given in Sec.~\ref{subsec: distance}.

The value $|\cdot\rangle_{a_1}$ of the ancilla qubits  is not relevant at this stage of the calculation, but we cannot discard it as it will be useful later.

Once the state is prepared, according to Grover's algorithm and its FPS counterpart, now we construct the oracle operator that applies a negative phase to the states where $d_{ij}\leq h$.

Here resides the novelty of this work: we do not know what is our target state $|\psi_T\rangle$, so we do not aim to build the oracle operator as described in Eq.~\eqref{eq:oracle_operator}, nevertheless we build the oracle by constructing a diagonal operator which is able to invert the sign of all the possible solutions.

\begin{equation}
    \label{eq: oracle operator - 1 excluded}
    \hat{O}(h) = \text{diag}\Bigl(\underbrace{-1,\dots,-1}_{h\text{ times}}, \underbrace{1,\dots,1}_{2^{(q_1+1)}-h\text{ times}}\Bigr).
    \end{equation}

\noindent When applied to $|\psi_1\rangle$ this yields

\begin{align}\label{eq:psi_2}
    |\psi_2\rangle&=(\mathbb{I}_{2q_0+q_1}\otimes\hat{O}(h)\otimes \mathbb{I}_{a_1})|\psi_1\rangle \\
    &=\nonumber
     \sqrt{\frac{N-M}{N}}\sum_{d_{ij}>h}|i\rangle_{q_0}|x_i\rangle_{q_1}|j\rangle_{q_0}|d_{ij}\rangle_{q_1}|\cdot\rangle_{a_1}
    \\
    &- \nonumber
    \sqrt{\frac{M}{N}}\sum_{d_{ij}\leq h}|i\rangle_{q_0}|x_i\rangle_{q_1}|j\rangle_{q_0}|d_{ij}\rangle_{q_1}|\cdot\rangle_{a_1}\nonumber\\
    & \nonumber
    =
    \cos{\theta}|\psi^\perp\rangle-\sin\theta|\psi_T\rangle,
\end{align}

\noindent where $M$ is the number of neighboring pairs and in the last line of Eq.~\eqref{eq:psi_2} we restored the convention that uses $|\psi^\perp\rangle,|\psi_T\rangle$ and the relative angle $\theta.$

\begin{figure*}[t]
    \centering
    \resizebox{.75\linewidth}{!}{
    \begin{quantikz}[row sep=0.7cm, column sep=0.7cm] 
    \lstick{$|0\rangle_{a_0}$} & \qw & \qw & \slice{$|\psi_0\rangle$}  & \gate[1]{\hat R_y(\alpha)} \slice{$|\psi_1\rangle$}         & \ctrl{4}\slice{$|\psi_2\rangle$}  &\gate[1]{\hat R^\dagger_y(\alpha)}&\ctrl{1} &\qw \slice{$|\psi_3\rangle$}  &\meter{}\\
    \lstick{$|0\rangle_{q_0}$} & \qw\gategroup[5,steps=4,style={fill=blue!20, inner xsep=2pt},background,label style={label
position=below,anchor=north, yshift=-0.2cm}]{$\hat U_1$} & \gate[2]{\text{PREP}}
    &\qw&\qw&\qw&\gate[5,style = {fill=blue!20}]{\hat U_1^{\dagger}}&\gate[2]{\bar Z}&\gate[5,style = {fill=blue!20}]{\hat U_1}&\qw \\
    \lstick{$|0\rangle_{q_0}$} & \permute{2,1} & \qw                   &\permute{2,1}& \qw                        & \qw   & \qw &\qw&\qw&\qw \\
    \lstick{$|0\rangle_{q_1}$} & \qw & \gate[2]{\text{PREP}} & \qw         & \gate[3]{\hat D}           & \qw             & \qw &\qw&\qw&\qw \\
    \lstick{$|0\rangle_{q_1}$} & \qw & \qw                   & \qw         & \qw                        & \gate[2]{\hat O}& \qw &\qw&\qw&\qw \\
    \lstick{$|0\rangle_{a_1}$} & \qw & \qw                   &\qw          & \qw                        & \qw             & \qw &\qw&\qw&\qw 
    \end{quantikz}
    }
    \caption{\label{fig:QFRANS} The circuit for the QFRANS algorithm, which has to be repeated until the state measured in the register $a_0$ is $|0\rangle$.}
    \label{fig: circuit explicit}
\end{figure*}
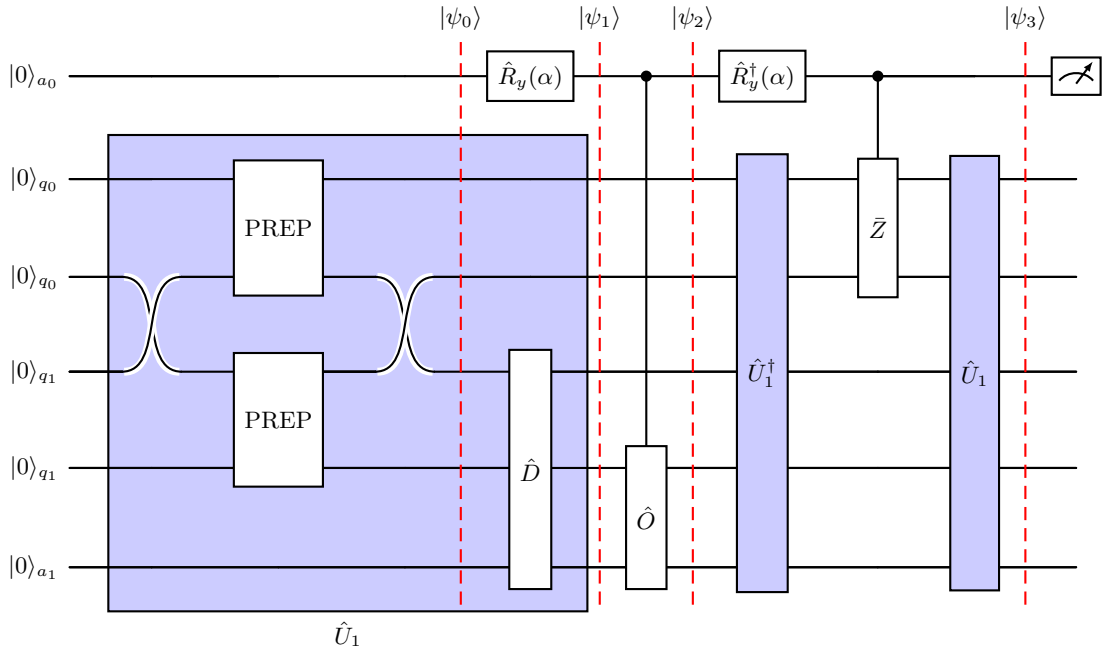

Finally we build the reflection over the initial state $\ket{\psi_1}$, $\hat{R_1}$, similar to Eq.~\eqref{eq:reflection_operator} as

\begin{equation}
    \label{eq: reflection over initial state}
    \hat{R}_1 = 2|\psi_1\rangle\langle\psi_1|-\mathbb{I} = \hat U_1(2|0\rangle\langle0|-\mathbb{I})\hat{U_1}^\dagger= \hat{U_1}\bar Z\hat{U_1}^\dagger,
\end{equation}

\noindent where we defined $\hat U_1 = \hat{D}\hat{U}$, and $\bar Z$ as the expanded $\hat Z$ operation which inverts the sign of the qubits different from $|0\rangle$. 
The application of $\hat R_1$ to the quantum state leads to $|\psi_3\rangle = \hat{R}_1 |\psi_2\rangle$.
Upon this premises, we are able to build the QFRANS algorithm as depicted in Fig.~\ref{fig:QFRANS}, which is analogue to the FPS algorithm, with an explicit state preparation for the FRANS problem, and a modified oracle operator. 
Although those differences, the result can be written as in Eq.~\eqref{eq:FPS_results}, with

\begin{align}
|\Psi_3\rangle &= 
-\sqrt{p_{k+1}} \, \sqrt{\frac{M}{N}} \sum_{d_{ij} \leq h} \left(  \ket{0}_{a_0}|i\rangle_{q_0}|x_i\rangle_{q_1}|j\rangle_{q_0}|d_{ij}\rangle_{q_1}|\cdot\rangle_{a_1} \right)
\\& \nonumber
+ \sqrt{1-p_{k+1}} \, \ket{1}_{a_0} \big{[}(c_k\cos2\theta - s_k\cos\alpha_k\sin2\theta) 
 \\ & \nonumber\sqrt{\frac{N-M}{N}}  \sum_{d_{ij} > h} \left(|i\rangle_{q_0}|x_i\rangle_{q_1}|j\rangle_{q_0}|d_{ij}\rangle_{q_1}|\cdot\rangle_{a_1} \right)
\\& 
+(c_k\sin2\theta  + s_k\cos\alpha_k\cos2\theta) 
\\ & \nonumber\sqrt{\frac{M}{N}}\sum_{d_{ij} \leq h} \left( |i\rangle_{q_0}|x_i\rangle_{q_1}|j\rangle_{q_0}|d_{ij}\rangle_{q_1}|\cdot\rangle_{a_1} \right)
\big{]},
\end{align}

In the remaining of this section we analyze each of the components of the algorithm separately.
\subsection{State preparation}
Here we describe the $\hat U_1$ operator needed to encode and process the pieces of information available in the database.

The fundamental premise of our algorithm relies on establishing a correspondence between dataset elements $\mathrm{X_i}$ and their integer representations $x_i$. When working with floating-point precision, this correspondence can always be achieved by defining a discretization grid based on the numerical precision of the computing system. More sophisticated grid selection schemes can be tailored to specific problem requirements and search radius $\xi$ to optimize performance.

To identify and mark the target states that satisfy the proximity condition in Eq.~\eqref{eq: oracle--operator}, we must first determine the integer representation of the chosen distance threshold $\xi$. This conversion is feasible precisely because we have established a bit-encoding scheme that associates each dataset point with a unique integer value.

A critical requirement of our approach is that the input dataset must be evenly spaced, or equivalently, that the data space can be accurately described using a regular grid structure. Under this constraint, we can interpret the integer representation as the number of discrete distance units required to span the threshold distance $\xi$. For illustration, consider the one-dimensional case where this relationship is expressed as:

\begin{equation}
    \label{eq: integer rappr of xi}
    h = \left\lceil \frac{\xi}{\Delta x} \right\rceil \,,
\end{equation}

where $\Delta x$ represents the grid spacing. 
This discretization procedure effectively transforms the original continuous proximity problem into an equivalent integer formulation:
\begin{equation}
    \label{eq: integer representation of the neighbour problem}
    d(x_i, x_j) \leq h \,, \quad  x_i, x_j, h \in \mathbb{N} \,.
\end{equation}
Hence, We assume that both the label and the position of each particle are described by integers. 
Note that the ceiling function in Eq.~\eqref{eq: integer rappr of xi} is suitable to a proximity condition that uses the inclusive inequality as in Eq.~\eqref{eq: integer representation of the neighbour problem}, whereas, for applications requiring a strict inequality, the floor function should favorable.

\subsubsection{The $\text{PREP}$ operator}\label{subsec: PREP}
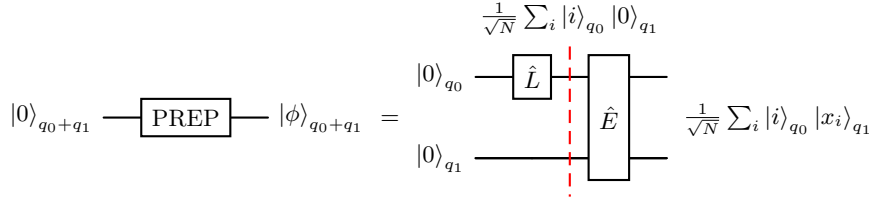
\begin{figure}[t]
    \centering
    \begin{quantikz}
        \lstick{$\ket{0}_{q_0+q_1}$} & \gate{\text{PREP}} & \rstick{$\ket{\phi}_{q_0+q_1}$}
    \end{quantikz} $=$
    \begin{quantikz}
        \lstick{$\ket{0}_{q_0}$} &\gate{\hat L}\slice{$\frac{1}{\sqrt{N}}\sum_i\ket{i}_{q_0}\ket{0}_{q_1}$} & \gate[2]{\hat E} & \\
        \lstick{$\ket{0}_{q_1}$} &\qw & \qw & \qw
    \end{quantikz} $\frac{1}{\sqrt{N}}\sum_i\ket{i}_{q_0}\ket{x_i}_{q_1}$
    
    \caption{A possible implementation of the $\text{PREP}$ operator, where $\hat L$ creates a balanced superposition of the labels of the particles and $\hat E$ assigns to each particle its position.}
    \label{fig:prep}
\end{figure}

In the following we treat $\text{PREP}$ as a black-box operator that is able to prepare the state $\ket{\phi}$ as defined in Eq.~\eqref{eq:PREP_operator}. 
The construction of $\text{PREP}$ is often disregarded but it is actually one of the hardest element to deal with. 
The quantum superposition of $N$ states can be obtained by employing a circuit with depth $\mathcal{O}(N)$, thus exponential with the number of qubits $q_0$ of the label register~\cite{malvetti_quantum_2021,mozafari_efficient_2021}. 
This can be interpreted as the sequential implementation of each particle directly on the quantum circuit, with the $\log N$ ancilla qubits facilitating the decomposition of multi-controlled gates.
Note that by using the optimized algorithm of Ref.~\cite{malvetti_quantum_2021}, we can implement $\text{PREP}$ as two separate circuit, as shown in Fig.~\ref{fig:prep}. The first operator, $\hat L$, defines the balanced superposition of $N$ elements in the register with $q_0$ qubits $\hat L |0\rangle_{q_0} = \frac{1}{\sqrt{N}}\sum_{i=0}^{N-1}|i\rangle$.
The second operator, $\hat E$, assigns to each particle its own position, $\hat E|i\rangle_{q_0}|0\rangle_{q_1} = |i\rangle_{q_0}|x_i\rangle_{q_1}$.

\noindent This implementation of $\text{PREP}$ allows us to use a reduced $\hat{R}_1$ operator, as we will see later.

One can drastically reduce the depth of the circuit by increasing its width, i.e. by employing ancilla qubits~\cite{zhang_low-depth_2021,zhang_quantum_2022,zhang_circuit_2024}, an approach that culminates with a circuit that has width$\sim\mathcal{O}(N)$ while keeping the depth$\sim\mathcal{O}(\log N)$.
This constructs a register of $N$ ancilla qubits containing the information for initial state preparation, effectively forming a database. Access to this repository scales logarithmically with the number of data points $N$, enabled by a tree-based data access structure. In our case, since the initial state has nearly uniform coefficients, the classical overhead for amplitude preparation is both constant and efficient.
In general, different algorithms play with the tradeoff between depth and number of ancillas~\cite{zhang_quantum_2022}, while keeping the product depth$\times$width about constant. 
In order to build a functioning $\text{PREP}$ operator, one should adapt the depth and the width to the specifics of the quantum computer available.  
Neither of the approaches at the extremes of this trade-off require all-to-all qubit connectivity; consequently, we do not expect connectivity constraints to significantly impact the algorithm's scaling.

Conversely, if the position of the particles follows some structure, that is it can be described by a mathematical function, we can use matrix access oracles with depth$\sim\mathcal{O}(\text{poly}\log N)$~\cite{zhang_circuit_2024,camps_explicit_2023,sanavio_carleman-lattice-boltzmann_2025}.

    \begin{figure}
        \centering
        \resizebox{\columnwidth}{!}{
        \begin{quantikz}
            \lstick{$\ket{|d_{ij}|}$} &\qwbundle{} &\gate[3]{\widehat{-1}} &\gate[4]{\text{COMP}} &\qw &\gate[4]{\text{COMP}^\dagger} &\gate[3]{\widehat{+1}} &\qw
            \\
            \lstick{$\text{sign}$} &\qw &\qw &\qw &\qw &\qw &\qw &\qw
            \\
            \lstick{$\ket{0}_{a_1}$} &\qwbundle{} & & &\qw & & &\qw
            \\
            \lstick{$\text{target}$} &\qw &\qw & &\gate{Z} & &\qw &\qw
        \end{quantikz}
        }
        \caption{Quantum Oracle. The circuit realizes the inverse oracle $-\hat{O}$ defined in Equation~\eqref{eq: oracle--operator}. It is important for the circuit to work correctly that the sign qubit is the most significant qubit. The target register stores the comparison result $d_{ij} \geq h$.  The operators $\widehat{\pm 1}$ respectively add and subtract unity to the distance value; they can be implemented with a logarithmic depth quantum adder~\cite{logarithmic_depth_quantum_adder}.
        Removing the incrementer $\widehat{+1}$ and decrementer $\widehat{-1}$ operators recovers the expression in Equation~\eqref{eq: oracle operator - 1 excluded}. The original oracle $\hat{O}$ can be recovered by replacing the $Z$ gate with $-Z$.}
        \label{fig: oracle implementation}
    \end{figure}

\subsubsection{The distance operator $\hat{D}$} \label{subsec: distance}
In a generic scenario, the distance operator $\hat{D}$ is applied on a combination of two identical states superpositions $\ket{\phi}\ket{\phi}$ 
\begin{equation}
    \label{eq: initial state}
    \hat{D} \sum_{i,j}\ket{x_i}_{q_1}\ket{x_j}_{q_1}|0\rangle_{a_1} = \sum_{i,j}\ket{x_i}_{q_1}\ket{d_{ij}}_{q_1+1} \ket{0}_{a_1-1} \, ,    
\end{equation}
where $\ket{d_{ij}}_{q_1+1}$ is the state encoding the signed distance between the particle $x_i$ and $x_j$. The extra qubit $q_1+1$ is borrowed from the ancillary register and used to represent the sign of the distance. 

Consider a three-dimensional Euclidean space with $L_2$ distance
${d(r_i, r_j) = \sqrt{(x_i-x_j)^2 + (y_i-y_j)^2 + (z_i-z_j)^2}}$. 
The integer FRANS problem in Eq.~\eqref{eq: integer representation of the neighbour problem} reduces to identifying particle pairs contained within a sphere of radius $h$. 
However, in certain scenarios, such as cosmological simulations~\cite{Gadget}, it is sufficient to determine whether particles fall within a cube of side length $h$ (Chebyshev distance). 
This approach decomposes the problem into three independent one-dimensional searches along each Cartesian axis, where the distance equals the absolute difference of the coordinates.
\begin{equation}
    \label{eq: cubic distance mapping}
    d(r_i, r_j) \leq \epsilon \mapsto |x_i - x_j| \leq h \,\wedge\, |y_i - y_j| \leq h \,\wedge\, |z_i - z_j| \leq h \,.
\end{equation}

Henceforth, we focus on the one-dimensional case, as the problem reduces to this simpler scenario when employing the Chebyshev distance. 

Before delving into details regarding the implementation of the absolute difference, we must explain how negative integers are represented in binary. Negative integer values are represented in binary form using two's complement: the binary representation of the integer is negated and $1$ is added, with the most significant bit (MSB) encoding the sign. For example, consider the integer $10$ with binary representation $01010$. We first negate the value to obtain $10101$, then add $1$ to yield $10110$, the binary representation of $-10$. To recover the absolute value of a negative integer, we apply two's complement again: negate the bits to obtain $01001$, then add $1$ to recover $01010$, corresponding to $10$. Thus, obtaining the absolute value requires negation followed by adding $1$ when the distance is negative.

In a quantum circuit, the difference between two binary values $a$ and $b$ is computed using the inverse Ripple Carry Adder~\cite{cuccaro_new_2004}, which requires $\mathcal{O}(q_1)$ CNOT gates and only 2 ancilla qubits: one for the carry of the subtraction and another as a clean ancillary qubit. When $a > b$, the carry qubit remains in state $|0\rangle$ and yields the correct result. When $a < b$, the carry qubit flips to $|1\rangle$, producing the two's complement representation of $a-b$. We thus treat the carry bit as the sign qubit and effectively as part of the distance.

To obtain the absolute value, when the sign qubit is $\ket{1}$, the incrementer $\widehat{+1}$ is applied to the distance register, performing the addition of unity: $\widehat{+1}\ket{d_{ij}}=\ket{d_{ij}+1}$. This controlled incrementer can be implemented with circuit depth $\mathcal{O}(q_1)$~\cite{egretta_thula, logarithmic_depth_quantum_adder}. Subsequently, the value is negated via a series of $q_1$ $CNOT$ gates, each controlled by the sign qubit and acting on $\ket{d_{ij}+1}$.

However, instead of using the absolute difference between particles, we propose to work with the signed difference. The negative values will be then disregarded as non-target states by an appropriate choice of the oracle; this will be explained in Sec~\ref{sec:IVB}. This operation is considerably simpler, requiring only the inverse adder and avoiding the controlled operations. Moreover, the number of target states is reduced by half, as each pair is counted only once.
\\
Summing up, the choice of the Chebyshev distance over the $L_2$ norm reduces the overall depth of the distance operator $\hat{D}$, as it allows the use of signed differences along each spatial dimension. In this context, $\hat{D}$ comprises three inverse adder operators, one acting along each spatial coordinate. This eliminates the squaring and summing operations required by the Euclidean distance~\cite{quantum_for_many_body}, thereby reducing the overall complexity of the distance operator.
\subsection{The oracle $\hat{O}$}\label{sec:IVB}
    The oracle presented in Equation~\eqref{eq: oracle operator - 1 excluded} represents the most naive implementation, as it incorporates elements with zero distances. This becomes particularly valuable when different particles are mapped to identical positions due to finite precision constraints. It is also the cheapest option in terms of ancilla qubits, number of CNOT and circuit depth.
    
    The most direct approach to implementing the oracle $\hat{O}$ employs the diagonal decomposition method described in Reference~\cite{diagonal_gate_Shende_2006}. However, this approach exhibits unfavorable scaling properties, with both the number of CNOT gates and circuit depth scaling as $\mathcal{O}(2^{q_1+1})$, rendering it impractical for large-scale implementations.

    An alternative approach would be to implement the oracle as a series of multi-controlled $Z$ gates (MCZ). In this scenario, we would require $\mathcal{O}(h)$ MCZ gates. Using the approach described in~\cite{vale_circuit_2024, zindorf_efficient_2025} the global scaling of the oracle would be $\mathcal{O}(hq_1)$.
    
    We can achieve a substantial improvement, reducing the complexity to $\mathcal{O}(q_1)$, by exploiting the inherent structure of our problem. In fact, we propose an oracle implementation that utilizes an integer comparator circuit (COMP), which requires $q_1-1$ clean ancilla qubits and one additional qubit to store the comparison result $d_{ij} \geq h$. When this condition is satisfied, the target qubit is found in state $\ket{1}$. The comparator's operation can be formally expressed as $\text{COMP}\left( \sum_{ij} \ket{d_{ij}}_{q_1+1}\ket{0}\right) = \sum_{ij} \ket{d_{ij}}_{q_1+1}\ket{d_{ij} \geq h}$. Choosing the most significant qubit as the sign bit, we set up a natural filtering mechanism. By adopting this choice, when we have a negative distance $d_{ij}$, the sign qubit is set to $\ket{1}$, which automatically makes the entire numerical value at least $2^{q_1}$. Since all negative values are now greater than $2^{q_1}-1$, the comparator naturally excludes them without any additional circuit. This elegant consequence of our sign bit placement simplifies the overall implementation.
    
    To optimize circuit complexity, we implement the inverse oracle $-\hat{O}$ rather than the oracle defined in Equation~\eqref{eq: oracle operator - 1 excluded}. This is achieved through a phase kickback mechanism: we first apply the quantum comparator to the distance register, then apply a $Z$ gate to the comparator's target qubit, and finally reset the target qubit to $\ket{0}$ by applying $\text{COMP}^\dagger$. This procedure introduces a negative phase to all elements whose distance satisfies $d_{ij} \geq h$. Using this approach the Oracle requires only $q_1$ clean ancilla qubits.

    However, when one is faced with large datasets, e.g, $N=10^9$ particles, we suggest to employ a slightly different oracle
    \begin{equation}
    \label{eq: oracle--operator}
    \hat{O}(h) = \text{diag}\Bigl(1, \underbrace{-1,\dots,-1}_{h-1\text{ times}}, \underbrace{1,\dots,1}_{2^{(q_1+1)}-h\text{ times}}\Bigr).
    \end{equation}
    where we introduced a modification to exclude zero-distance values $\ket{0}$. This eliminates comparisons between identical data points, reducing the number of solutions by $N$. In practical applications, e.g, cosmological simulations where $N \sim 10^{10}$, this optimization provides significant computational advantages. 
    
    Figure~\ref{fig: oracle implementation} illustrates the quantum circuit implementation of $-\hat{O}$ from Equation~\eqref{eq: oracle--operator}. Zero values are excluded from the marked states by replacing the standard comparator COMP with $\text{COMP} \, (\widehat{-1})$, where $\widehat{-1}$ represents the inverse quantum incrementer based on a logarithmic depth quantum adder~\cite{logarithmic_depth_quantum_adder}.
    To maintain linear scaling in both CNOT count and circuit depth with respect to $q_1$, this modified approach requires $2q_1 - \log_2q_1$ clean ancilla qubits~\cite{egretta_thula}, which allows us to use the same ancilla qubits for both the comparator and the incrementer. 

\begin{figure*}[t]
        \centering
        \includegraphics[width=1\linewidth]{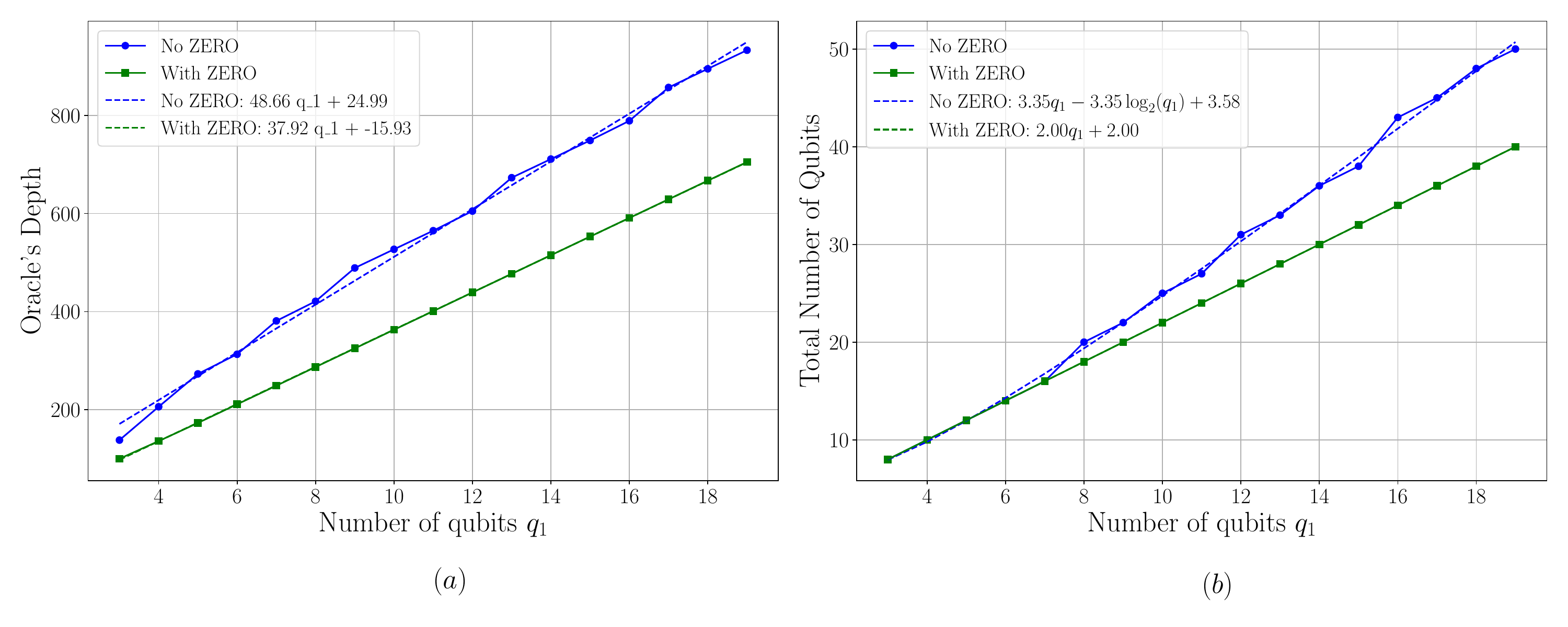}
        \caption{Depth $(a)$ and total number of qubits $(b)$ as a function of the number of qubits $q_1$. The green line refers to the oracle built using the comparator for the case including Zero as in Eq.~\eqref{eq: oracle--operator}); the blue line to the case without zero as in Eq.~\eqref{eq: oracle operator - 1 excluded}.}
        \label{fig: comp comparison}
    \end{figure*}

    In Fig.~\ref{fig: comp comparison} we show a comparison of the depth and number of ancilla qubits required to implement both version of the oracle. We have to consider also that the fewer the number of target states $M$, the harder is to find them. Specifically the number of required repetition is expected to scale as $\mathcal{O}(N/\sqrt{M})$ (cf. Sec~\ref{sec:III}, as $N^2$ is the number of all the possible pairs of the dataset with $N$ elements.

    \subsection{The reflection}\label{sec:IVC}
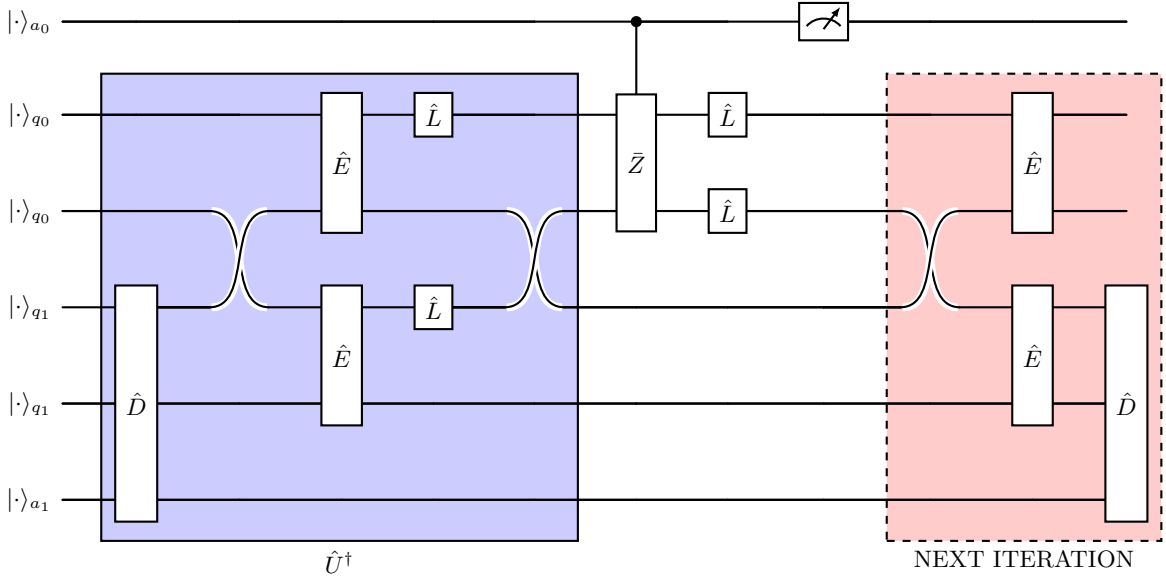
\begin{figure*}
    \centering
    \resizebox{.8\linewidth}{!}{
    \begin{quantikz}[row sep=0.7cm, column sep=0.7cm] 
    \lstick{$|\cdot\rangle_{a_0}$} 
    &\qw     & \qw   & \qw  &\qw  &\qw   &\ctrl{1}   &\qw    &\meter{}   &\qw    &\qw   &\qw
    \\
    \lstick{$|\cdot\rangle_{q_0}$} &\qw\gategroup[5,steps=5,style={fill=blue!20, inner xsep=2pt},background,label style={label
position=below,anchor=north, yshift=-0.2cm}]{$\hat U^\dagger$} 
    &\qw    &\gate[2]{\hat{E}}      &\gate{\hat{L}} &\qw    &\gate[2]{\bar{Z}}  &\gate{\hat{L}}    &\qw
    &\qw \gategroup[5,steps=3, style={dashed, fill=red!20, inner xsep=2pt},background,label style={label
position=below,anchor=north, yshift=-0.2cm}]{NEXT ITERATION}  &\gate[2]{\hat{E}} &\qw
    \\
    \lstick{$|\cdot\rangle_{q_0}$}  &\qw &\permute{2,1}     & \qw    &\qw   &\permute{2,1}          &\qw   &\gate{\hat{L}}    &\qw    &\permute{2,1}       &\qw  &\qw
    \\
    \lstick{$|\cdot\rangle_{q_1}$} &\gate[3]{\hat{D}}   &\qw    &\gate[2]{\hat{E}}  &\gate{\hat{L}}     &\qw      &\qw   &\qw    &\qw   &\qw &\gate[2]{\hat{E}}  &\gate[3]{\hat{D}} 
    \\
    \lstick{$|\cdot\rangle_{q_1}$} &\qw     &\qw      &\qw         & \qw         & \qw        &\qw     & \qw         & \qw         & \qw         & \qw         & \qw         
    \\
    \lstick{$|\cdot\rangle_{a_1}$} &\qw     &\qw     &\qw      &\qw          & \qw        &\qw     & \qw         & \qw         & \qw           & \qw         & \qw            
    \end{quantikz}
    }
    \caption{Explicit representation of the $\hat{R_0}$ operator. The part included by the red-dashed line is applied only if the ancilla measurement is $\ket{1}$.}
    \label{fig: R0}
\end{figure*}

The final step of the QFRANS algorithm requires performing an inversion around the initial state $\ket{\psi_1}$. As demonstrated in Eq.~\eqref{eq: reflection over initial state} and illustrated in Figs.~\ref{fig: circuit explicit}~\ref{fig: R0}, this can be achieved through the implementation of a single operator $\bar{Z}$. \\
To maintain consistency with our circuit optimization strategy introduced in Section~\ref{sec:IVB}, where we implemented the negative oracle $-\hat{O}$ rather than the positive version, we adopt the same approach here by constructing $-\hat{R}_1$ instead of $\hat{R}_1$. This design choice ensures that the accumulated negative phases from both the oracle and reflection operators cancel out, ultimately producing results that are identical to the standard implementation presented in Section~\ref{sec:IV}. Under this modified framework, the required operator becomes $-\bar{Z} = \mathbb{I}-2\ket{0}\bra{0}$, which can be efficiently realized as a multi-controlled $Z$ gate that activates when all control qubits are in the $\ket{0}$ state.

Figure~\ref{fig: R0} presents an explicit representation of the $R_0$ implementation. As established in Section~\ref{subsec: PREP} and illustrated in Figure~\ref{fig:prep}, $\text{PREP}$ can be factorized as the product of label creation operator $\hat{L}$ and element creation operator $\hat{E}$. Consequently, following application of $\hat{\bar{Z}}$, the complete $\hat{U}$ operator need not be applied in all cases. 
When readout measurements target only the labels of dataset elements, applying $\hat{L}$ suffices to create the particle labels, whereas $\hat{U}$ would reconstruct a superposition of particle elements and distances. However, when the ancilla measurement yields $\ket{1}$, correct algorithmic behavior requires the full $\hat{U}$ operator. This is achieved by applying the remaining components $\hat{E}$ and $\hat{D}$ to the appropriate registers.

The scaling of $\hat{R_0}$ is dominated by the $\text{PREP}$ operator. In fact,~\cite{zindorf_efficient_2025} shows that is possible to implement a multi-controlled $X$ gate using only $12n + \mathcal{O}(1)$ qubits and one dirty ancilla, where in the QFRANS scenario $n = 2q_0$. Consequentially, even with an advantageous decomposition of the $MCZ$, the implementation f $\hat{R_0}$ remains the bottleneck of the algorithm, as the depth and complexity scales in the same manner as $\text{PREP}$.

\subsection{The readout measurement}
The protocols described in Sec.~\ref{sec:IVB} and Sec.~\ref{sec:IVC} have to be repeated until we measure the state $|0\rangle_{a_0}$ on the ancilla qubit. 
When this happens, supposing an error-free quantum hardware, the system has collapsed to the target state, an uniform superposition of the $M$ solutions. 

We propose to use the Bayes rule to estimate the value $M$ representing the number of solution states. We treat $M$ as a random variable and assume, under the hypothesis $N \gg M$, a Poisson prior probability distribution 
\begin{equation}\label{eq:prior}
    p(M) = \frac{\mu^{M}e^{-\mu}}{M \, !} \,,
\end{equation}
with the mean number of solutions' pairs $\mu = \sum_M \, M\,p(M)$. 
The initial value for $\mu$ can be derived from previous simulation results or a priori knowledge of the problem (e.g., statistical information from classical simulations). In the absence of such prior knowledge, we set $\mu$ to the expected number of neighboring pairs in a uniform particle distribution $\left(\frac{2h}{L}\right)^dN^2$.
    
The probability that in a FRANS problems with $M$ solutions, the QFRANS algorithm converges to the target state after $m$ queries is given by the probability of failure in the first $m-1$, combined with success at the $m$-th one: $p_0(m|M) =p_m(M)\prod_{i=1}^{m-1}(1-p_i(M))$. Here $p_i(M)$ and $p_m(M)$ are functions of $M$, and their expression is  given in Eq.~\eqref{eq:instantaneous_prob}. The dependence on $M$ is hidden in the $\theta$ coefficients of Eq.~\eqref{eq:sincos_FPS}, explicitly $\theta = \arcsin \sqrt{M/N}$.
When after $m$ oracle's queries the ancilla is measured in state $\ket{0}$, the algorithm has converged to the solution. Following the readout measurements,  the prior distribution is updated accordingly to the Bayes rules
\\
\begin{equation}
    \label{eq: bayes probability}
    p(M|m) = \frac{p_0(m|M)\times p(M)}{\sum_M' p_{0}(m|M')p(M')} \,.
\end{equation}

Using the probability in Eq.~\eqref{eq: bayes probability} is possible to obtain a better estimate of the number of solutions $\mu = \sum_M M p(M|m)$. 
This updated value provides a principled basis for establishing an upper bound on the number of oracle queries. Since the algorithm is expected to converge in $\mathcal{O}(\sqrt{N^2 / M})$ iterations, and  $\mu$ serves as an estimate of $M$, we can set the iteration limit to $\mathcal{O}(\sqrt{N^2 / \mu})$.

To determine when the QFRANS sampling procedure should stop, we adopt a Bayesian stopping criterion based on the probability of observing new solutions.
Using the updated posterior distribution in Eq.~\eqref{eq: bayes probability} the  probability that the next sample will reveal a previously unseen solution:
\begin{equation} \label{eq: stoppage criterion}
P_{\mathrm{new}} = \mathbb{E}_{M\mid \text{data}}\left[1 - \frac{K}{M}\right] = \sum_{M=k}^{N} p(M,m) \left(1 - \frac{K}{M}\right),
\end{equation}
where $K$ is the number of distinct solutions observed thus far. This criterion directly reflects the operational goal of determining whether further QFRANS iterations are likely to yield new information. The algorithm stops when $P_{\mathrm{new}}$ falls below an arbitrary threshold: $P_{new}<\varepsilon$.

\section{Numerical Tests}\label{subsec: numerical tests}

\begin{figure*}
    \centering
    \includegraphics[width=.95\linewidth]{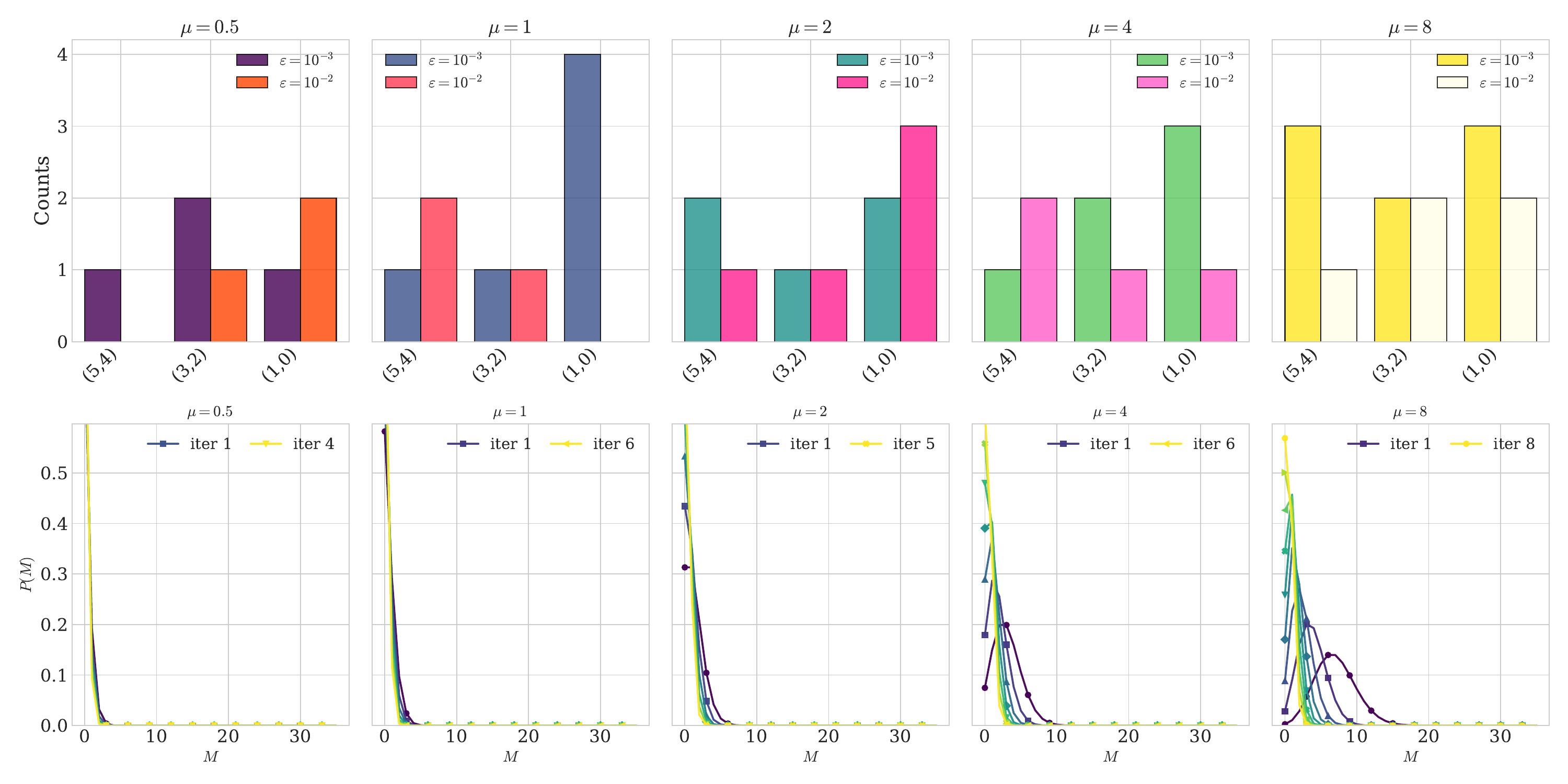}
    \caption{ Results from simulation with $N=36, M=3$ and $5$ maximum queries  allowed per run. The results are presented for different initial mean guesses $\mu$ in the prior distribution of Eq.~\eqref{eq:prior}. In the top row is shown the outcome of the readout, listing name of the particles on the x axis and their respective occurrences for different stopping tolerance $\varepsilon$. In the bottom row the evolution of the prior probability in Eq.~\eqref{eq: bayes probability}
 as a function of $M$., in the case where $\varepsilon=10^{-3}$.}
 \label{fig: hist and bayes M = 3}
\end{figure*}

\begin{figure*}
    \centering
    \includegraphics[width=.95\linewidth]{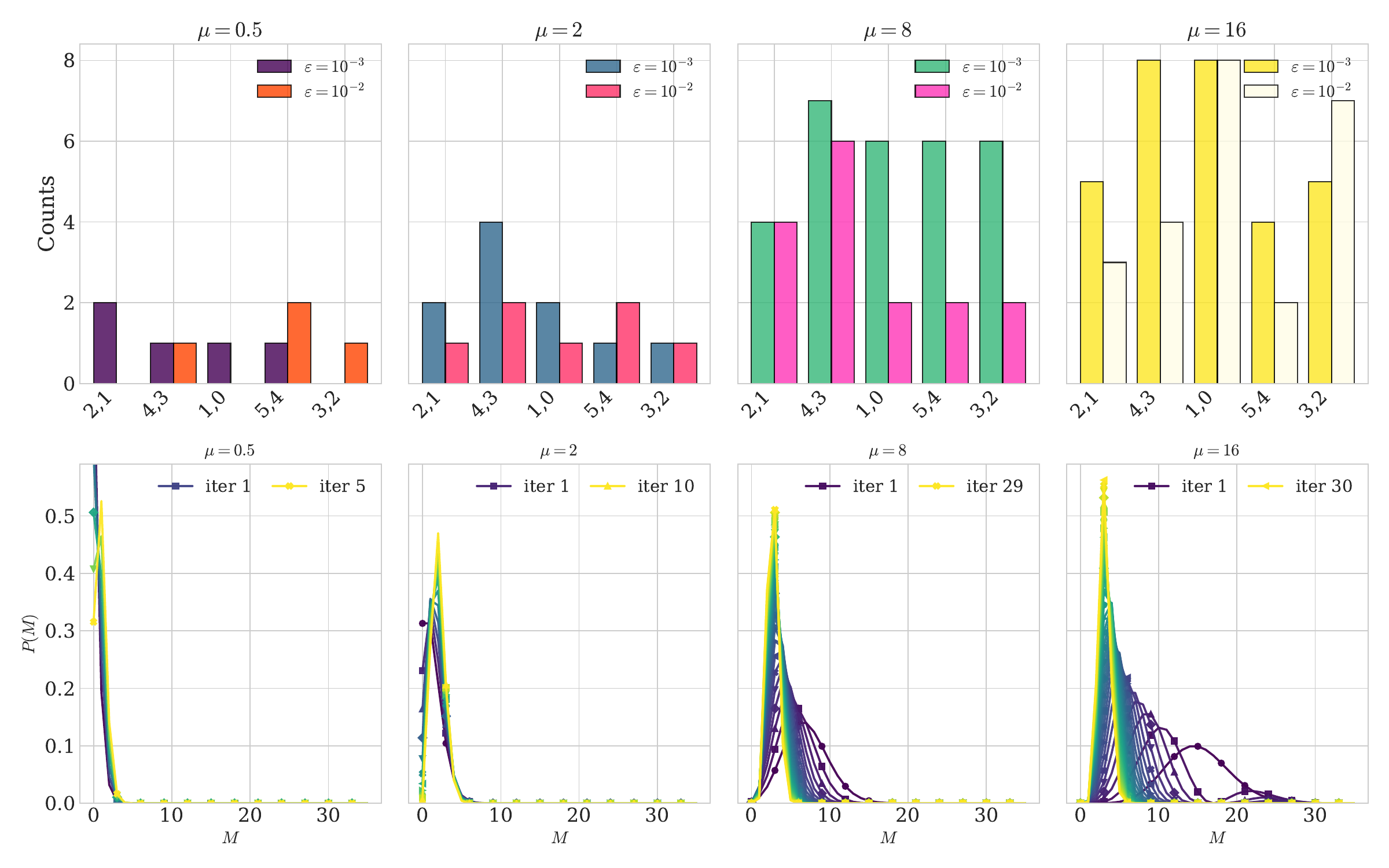}
     \caption{ Results from simulation with $N=36, M=5$ and $3$ maximum queries  allowed per run. The results are presented for different initial mean guesses $\mu$ in the prior distribution of Eq.~\eqref{eq:prior}. In the top row is shown the outcome of the readout, listing name of the particles on the x axis and their respective occurrences for different stopping tolerance $\varepsilon$. In the bottom row the evolution of the prior probability in Eq.~\eqref{eq: bayes probability}
 as a function of $M$., in the case where $\varepsilon=10^{-3}$. For both set of simulations the maximum number of allowed iteration in $30$.
 }
    \label{fig: hist and bayes M=5}
\end{figure*}

We have run simulations to verify the behavior of QFRANS and the stopping criterion under different conditions. We considered a 1 dimensional setup, where 6 particles are arranged in a box of dimension $L=8$: $\boldsymbol{x} = \{0; 1; 3; 4; 6; 7 \}$. We first choose an integer smoothing length $h=1$ such that $M = 3$ with neighboring result couple being $(x_0,x_1); (x_2,x_3); (x_4,x_5)$. 
Figure~\ref{fig: hist and bayes M = 3} illustrates the interplay between the prior mean $\mu$ and convergence tolerance $\varepsilon$ (Eq.~\eqref{eq: stoppage criterion}) in determining the probability of recovering all correct solutions. For a given $\varepsilon$, underestimating $\mu$ accelerates convergence at the risk of missing solution pairs, while overestimating $\mu$ ensures complete recovery but requires smaller $\varepsilon$ values and incurs higher computational costs through increased oracle queries. Our analysis reveals that $\varepsilon = 10^{-2}$ strikes an effective balance, guaranteeing robust convergence provided the initial mean is not severely underestimated. This behavior is consistent across both $M = 3$ and $M = 5$ cases, as shown in Fig.~\ref{fig: hist and bayes M=5}, although simulations with $\mu = 16$ reach the maximum iteration limit of 30 before achieving convergence.
\\
A key finding is that complete solution recovery remains possible even when the probability distribution $p(M)$ fails to converge to a sharp peak around the true value. This robustness arises directly from our convergence criterion (Eq.~\eqref{eq: stoppage criterion}), which validates solutions through an independent mechanism rather than relying solely on the estimated mean.

Moreover, to better assess how the choice of $\mu$ and $\varepsilon$ affects both the convergence rate and the number of solutions found, we ran a series of tests. Specifically, fixing $M = 5$, we performed nine simulations for each $(\mu, \varepsilon)$ configuration in order to compute the average number of iterations and the average number of solutions recovered. We repeated this procedure for two settings: one in which the maximum number of allowed queries was $3$, and another in which it was $5$.
\\
From the results reported respectively in Tabs.~\ref{tab: reps = 3} and \ref{tab: reps = 5}, we observe that restricting the number of queries tends to increase the likelihood of finding all solutions at the cost of slower convergence. Conversely, allowing too many queries can drastically reduce the probability of recovering all solutions under the same $(\mu,\varepsilon)$ configuration. This occurs because the probability in Eq.~\eqref{eq: bayes probability} depends on the likelihood of obtaining a measurement within $m$ queries: larger $m$ speeds up convergence but reduces accuracy, namely, the ability to recover all solutions.
\\
Moreover, recalling that the algorithm converges in $\mathcal{O}\!\left(\sqrt{N^2/M}\right)$ queries, we point out how in this case such value correspond to approximately $\sqrt{36/5}\simeq3$. This value thus represent s the ideal threshold for allowed queries. However, the true value of $M$ is not known a priori and only have its estimate or guess $\mu$ is available. Thus, to provide a more standardized and broadly applicable procedure, we repeated the simulations where the maximum number of queries adapted to the estimated mean value $\mu$, according to
\begin{equation}
    \label{eq: queries strategy}
    \left\lceil 1.5\,\sqrt{\frac{N^2}{\mu}} \right\rceil \, .    
\end{equation}
From the results presented in Tab.~\ref{tab: reps = adapt} is clear that the key requirement for finding all solutions is to overestimate the true number of solutions. The choice $\varepsilon = 10^{-2}$ appears to offer a good compromise between convergence speed and the possibility of finding all solutions, even when the overestimation is modest.

\begin{table*}
\centering

\caption{Mean number of iteration and found solutions as a function of the tolerance for the stopping criterion $\varepsilon$ and estimated number of solutions $\mu$. The results are obtained using $3$ maximum allowed queries and the means and errors are evaluated on a sample of $9$ data.}
\label{tab: reps = 3}
\setlength{\tabcolsep}{8pt}
\renewcommand{\arraystretch}{1.2}

\begin{tabular}{c ccc ccc ccc ccc}
\toprule
 & \multicolumn{2}{c}{$\varepsilon=10^{-1}$} 
 & \multicolumn{2}{c}{$\varepsilon=10^{-2}$}
 & \multicolumn{2}{c}{$\varepsilon=10^{-3}$}
 & \multicolumn{2}{c}{$\varepsilon=10^{-4}$} \\
\cmidrule(lr){2-3}\cmidrule(lr){4-5}\cmidrule(lr){6-7}\cmidrule(lr){8-9}
$\mu$ 
& Solution & Iteration 
& Solution & Iteration
& Solution & Iteration
& Solution & Iteration \\
\midrule
0.5 & 2.0 $\pm$ 0.0 & 2.0 $\pm$ 0.0 & 3.0 $\pm$ 0.0 & 3.6 $\pm$ 0.6 & 4.0 $\pm$ 0.0 & 7.3 $\pm$ 1.0 & 5.0 $\pm$ 0.0 & 12.6 $\pm$ 3.7 \\
2 & 3.0 $\pm$ 0.0 & 3.2 $\pm$ 0.3 & 4.3 $\pm$ 0.4 & 9.0 $\pm$ 1.5 & 5.0 $\pm$ 0.0 & 12.4 $\pm$ 2.2 & 5.0 $\pm$ 0.0 & 22.3 $\pm$ 0.77 \\
8 & 4.6 $\pm$ 0.4 & 10.7 $\pm$ 1.6 & 5.0 $\pm$ 0.0 & 16.3 $\pm$ 0.8 & 5.0 $\pm$ 0.0 & 29.0 $\pm$ 0.0 & 5.0 $\pm$ 0.0 & 30.0 $\pm$ 0.0 \\
16 & 5.0 $\pm$ 0.0 & 11.4 $\pm$ 1.0 & 5.0 $\pm$ 0.0 & 24.0 $\pm$ 0.0 & 5.0 $\pm$ 0.0 & 30.0 $\pm$ 0.0 & 5.0 $\pm$ 0.0 & 30.0 $\pm$ 0.00 \\
\bottomrule
\end{tabular}
\end{table*}

\begin{table*}
\centering
\caption{Mean number of iteration and found solutions as a function of the tolerance for the stopping criterion $\varepsilon$ and estimated number of solutions $\mu$. The results are obtained using $5$ maximum allowed queries and the means and errors are evaluated on a sample of $9$ data.}
\label{tab: reps = 5}
\setlength{\tabcolsep}{8pt}
\renewcommand{\arraystretch}{1.2}

\begin{tabular}{c ccc ccc ccc ccc}
\toprule
 & \multicolumn{2}{c}{$\varepsilon=10^{-1}$} 
 & \multicolumn{2}{c}{$\varepsilon=10^{-2}$}
 & \multicolumn{2}{c}{$\varepsilon=10^{-3}$}
 & \multicolumn{2}{c}{$\varepsilon=10^{-4}$} \\
\cmidrule(lr){2-3}\cmidrule(lr){4-5}\cmidrule(lr){6-7}\cmidrule(lr){8-9}
$\mu$ 
& Solution & Iteration 
& Solution & Iteration
& Solution & Iteration
& Solution & Iteration \\
\midrule
0.5 
& $1.0 \pm 0.0$ & $1.0 \pm 0.0$
& $2.0 \pm 0.0$ & $2.3 \pm 0.5$
& $2.9 \pm 0.3$ & $4.0 \pm 0.6$
& $3.0 \pm 0.0$ & $3.7 \pm 0.9$ \\
2 
& $2.0 \pm 0.0$ & $2.4 \pm 0.6$
& $2.9 \pm 0.3$ & $4.0 \pm 0.7$
& $3.1 \pm 0.3$ & $4.8 \pm 1.1$
& $3.4 \pm 0.4$ & $5.6 \pm 0.8$ \\
8 
& $2.7 \pm 0.4$ & $3.9 \pm 0.8$
& $3.2 \pm 0.3$ & $5.2 \pm 0.8$
& $3.6 \pm 0.4$ & $6.7 \pm 1.1$
& $4.4 \pm 0.4$ & $6.8 \pm 0.8$ \\
16 
& $3.7 \pm 0.4$ & $4.9 \pm 0.5$
& $3.8 \pm 0.5$ & $6.4 \pm 0.6$
& $4.6 \pm 0.4$ & $7.6 \pm 0.4$
& $4.4 \pm 0.6$ & $9.3 \pm 0.8$ \\
\bottomrule
\end{tabular}
\end{table*}

\begin{table*}
\centering
\caption{Mean number of iteration and found solutions as a function of the tolerance for the stopping criterion $\varepsilon$ and estimated number of solutions $\mu$. The results are obtained using an iterative maximum number of queries that adapts to the estimate of $\mu$. The means and errors are evaluated on a sample of $9$ data.}
\label{tab: reps = adapt}
\setlength{\tabcolsep}{8pt}
\renewcommand{\arraystretch}{1.2}

\begin{tabular}{c ccc ccc ccc ccc}
\toprule
 & \multicolumn{2}{c}{$\varepsilon=10^{-1}$} 
 & \multicolumn{2}{c}{$\varepsilon=10^{-2}$}
 & \multicolumn{2}{c}{$\varepsilon=10^{-3}$}
 & \multicolumn{2}{c}{$\varepsilon=10^{-4}$} \\
\cmidrule(lr){2-3}\cmidrule(lr){4-5}\cmidrule(lr){6-7}\cmidrule(lr){8-9}
$\mu$ 
& Solution & Iteration 
& Solution & Iteration
& Solution & Iteration
& Solution & Iteration \\
\midrule
0.5 & 1.0 $\pm$ 0.0 & 1.0 $\pm$ 0.0 & 1.0 $\pm$ 0.0 & 1.0 $\pm$ 0.0 & 2.0 $\pm$ 0.0 & 2.2 $\pm$ 0.3 & 2.0 $\pm$ 0.0 & 2.4 $\pm$ 0.6 \\
2 & 1.8 $\pm$ 0.3 & 2.0 $\pm$ 0.0 & 2.0 $\pm$ 0.0 & 2.0 $\pm$ 0.0 & 2.3 $\pm$ 0.3 & 3.1 $\pm$ 0.3 & 2.9 $\pm$ 0.3 & 3.6 $\pm$ 0.6 \\
8 & 3.3 $\pm$ 0.4 & 4.2 $\pm$ 0.3 & 3.9 $\pm$ 0.3 & 5.4 $\pm$ 0.4 & 4.1 $\pm$ 0.3 & 6.1 $\pm$ 0.3 & 3.9 $\pm$ 0.6 & 6.6 $\pm$ 0.4 \\
16 & 4.2 $\pm$ 0.5 & 7.3 $\pm$ 0.7 & 4.7 $\pm$ 0.4 & 9.2 $\pm$ 1.0 & 4.8 $\pm$ 0.3 & 10.3 $\pm$ 0.4 & 4.8 $\pm$ 0.3 & 11.4 $\pm$ 0.56 \\
\bottomrule
\end{tabular}
\end{table*}

\section{Analysis of noise resilience}\label{sec:V}

Today's quantum hardware is affected by noise and implementing any search algorithm in a scalable system is a significant challenge \cite{figgatt_complete_2017, Zhang_2022}. Only recently, a novel noise tolerant method has been developed to reduce the error threshold for Grover's search by optimizing the number of iterations \cite{PhysRevResearch.7.L012017}.

In this section we employ a simple model in order to analyze the effect of the noise on the result of the QFRANS algorithm assuming to have a quantum computer that can be characterized only by a bit-flip readout error. This assumption, though quite restrictive, can be used to characterize the noise resilience of the proposed algorithm. Therefore, we consider that the state, after successful measurement of the ancilla qubit onto $|1\rangle$, has collapsed into the correct target state $|\psi_T\rangle$ and that noise affects the algorithm only in the readout measurement part. We will see that this assumption does not change our proposed strategy which does not rely on error correction codes. 

The bit-flip channel noise flips the state of a qubit with a probability $1-p=\text{error rate}$, where $p$ is the probability of getting the correct result. This error can affect the two registers that are actually measured, both with $q_0$ qubits, which give the labels corresponding to the two particles. Starting from the states $\ket{i}_{q_0}, \ket{j}_{q_0}$, the overall probability of successfully measuring the exact indexes $i$ and $j$ is given by $p^{2 q_0}$. As a result, if we set a tolerance value $\texttt{TOL}$ for this probability, i.e. $p^{2q_0}\geq\texttt{TOL}$, the error rate must satisfy the following inequality:
\begin{equation}
     \text{error rate}\leq 1-\texttt{TOL} ^{\frac{1}{2q_0}}.
\end{equation}
Using reasonable values for the tolerance $\texttt{TOL}=0.99$ and the number of qubits $q_0=30$ to encode a large database of $N=2^{30}$ particles, the required error rate is in the order of $10^{-4}$. Within this simplified model, the proposed algorithm provides sufficiently accurate results within currently available error rates.
To improve the robustness of the proposed algorithm to noise, post-readout classical measurement techniques can be employed. For instance, if the full reflection operator is applied at the last iteration, measuring the state of the distance register $q_1$ provides a direct method for error detection: an error has occurred if the measured distance exceeds the integer smoothing length $h$. Moreover, a mismatch between the position register $q_1$ and the label of the first particle can also reveal erroneous circuit behavior, indicating that the readout should be rejected.

\section{Discussion and Conclusions}\label{sec:VI}
One of the main bottleneck in N-body methods is the FRANS subroutine. This is due to cache misses that arise when data that are close in space resides in distant memory locations. In this work we explored a possible solution using quantum computers. We emphasize that our goal was not to compete with state-of-the-art classical methods, but rather to introduce quantum algorithms aimed at finding alternative solutions to a class of problems that classically suffer from cache misses. We regard the present work as a preliminary step in a direction that may draw significant benefits from future developments of quantum computing. 

Building on fixed point amplitude amplification we developed QFRANS, a quantum algorithm that finds all the elements whose distance is less than a fixed radius $\xi$  with the use of an ancilla qubit that signals if the quantum state has reached the target.
We developed an explicit encoding of the data that, by leveraging quantum superposition, simultaneously evaluates all the $N^2$ distances between particles. This allows to recover with $\mathcal{O}(M \log M)$ measures and a single quantum circuit all the pairs of close neighbors, as opposed to classical FRANS algorithms that repeat the search process for each particle with the risk of incurring in cache miss.
One of the novelty of this work resides in the \textit{oracle}, which is efficiently implemented on a quantum circuit by using two comparators and with a depth scaling linearly with the number of qubits. We proposed two different versions of such oracle, that can be used in different scenarios.
We introduced the possibility of using the Chebyshev distance in place of the Euclidean one, bringing the complexity of the oracle to $\mathcal{O}(q_1)$ where $2^{q_1}$ is the number of grid points used to discretize one dimension.
Our algorithm is robust against bit-flip errors in the readout measurements process thanks to the encoding of data labels and distances that allows an easy-to-implement error detection.
We proposed and tested on a $1D$ simulation an efficient stopping criterion based on classical Bayes interference. 

We found that the probability and speed of recovering all solutions also depend on the maximum number of queries allowed. We propose an adaptive strategy that sets this threshold based on the estimation of $M$. Our observations show that with a tolerance value of $\varepsilon=10^{-2}$, this approach provides a good compromise between the number of iterations and the recovery of all neighboring particles, provided that $M$ is over-estimated.

Even though we employed state of the art algorithms, state preparation represents the major contribution to the circuit depth of our algorithm. For unstructured datasets, current methods face a fundamental depth-width trade-off: implementations requiring $\mathcal{O}(N)$ circuit depth utilize $\mathcal{O}(\log N)$ ancilla qubits, while those achieving $\mathcal{O}(\log N)$ depth demand $\mathcal{O}(N)$ ancilla qubits. 
Intermediate strategies exist between these extremes, though all maintain the constraint $\text{depth} \times \text{width} = \mathcal{O}(N \log N)$.

The fixed-point algorithm necessitates $\sqrt{N^2/M}$ queries to achieve the target state, which, when combined with the cost of implementing the oracle and diffusion operator, yields overall depth $\mathcal{O}(M^{-1/2} N^2)$ with $\mathcal{O}(\log N)$ ancilla qubits, or $\mathcal{O}((M \log M)^{1/2} N \log N)$ depth with $\mathcal{O}(N)$ ancilla qubits, depending on the encoding protocol. In addition we need to account for the readout requiring $\mathcal{O}(M\log M)$ measurements.
Although at first glance the computational complexity of QFRANS does not appear to offer substantial advantages over the best classical methods ($\mathcal{O}(N \log N)$), a couple of considerations should be weighed before drawing firm conclusions. The first concerns cache-misses in classical FRANS codes. This issue primarily slows down execution time because the volume of data the CPU must process is too large to be kept in cache and therefore resides in RAM. The classical codes are then heavily parallelized to reduce the amount of data streamed to the CPU, which shortens run times and lowers overall complexity.
\\
By analogy, the main obstacle to implementing QFRANS is the enormous cost of state preparation. In particular, choosing to work with $\mathcal{O}(N)$ ancilla qubits and depth $\mathcal{O}(\log N)$ is equivalent to having a particles' database with logarithmic access. If one were to parallelize the algorithm, the load on the QPU would decrease, making an implementation more plausible. Once again, QFRANS does not suffer from cache misses as there is no QRAM~\cite{qram} and the burden falls entirely on the QPU. This will be analogous to a classical cache-only implementation. Nevertheless, the limiting factor here is still the sheer amount of data, thus for the moment, QFRANS should be regarded as a proof of concept, as well as a cache only classical implementation.

In future research more efficient state preparation algorithms will be investigated. As anticipated, a more detailed study on domain decomposition strategies may offer computational advantages, though at the cost of requiring multiple circuit implementations. Additionally, detailed studies of specific applications, such as cosmological simulations or molecular dynamics as well as flows on dynamic porous media may be used to validate the algorithm's practical utility. Understanding error propagation and developing robust mitigation strategies will be considered in future studies.
QFRANS represents a promising foundation for quantum algorithms addressing neighbor search problems. This work suggests that quantum computing may offer genuine advantages in computational domains where classical methods face fundamental memory hierarchy limitations rather than purely algorithmic constraints.

\begin{acknowledgments}
We thank Simona Perotto, Filippo Marchetti, Stefano Borgani and Luigi Iapichino for valuable discussions.
LC and GM have benefited scientifically from the collaboration with the Spoke 10 INAF group.
LC, GM, CS and SS acknowledge financial support form the Italian National Center for HPC, Big Data and Quantum Computing (CN00000013).
\end{acknowledgments}

\bibliography{bibliography}

@book{nielsen_quantum_2010,
	title = {Quantum {Computation} and {Quantum} {Information}: 10th {Anniversary} {Edition}},
	isbn = {978-1-139-49548-6},
	url = {https://books.google.it/books?id=-s4DEy7o-a0C},
	publisher = {Cambridge University Press},
	author = {Nielsen, M.A. and Chuang, I.L.},
	year = {2010},
}

@INPROCEEDINGS{grover_fast_1996,
author = {Grover, Lov K.}, title = {A fast quantum mechanical algorithm for database search}, year = {1996}, isbn = {0897917855}, publisher = {Association for Computing Machinery}, address = {New York, NY, USA}, url = {https://doi.org/10.1145/237814.237866}, doi = {10.1145/237814.237866}, booktitle = {Proceedings of the Twenty-Eighth Annual ACM Symposium on Theory of Computing}, pages = {212–219}, numpages = {8}, location = {Philadelphia, Pennsylvania, USA}, series = {STOC '96}
}

@article{grover_fixed-point_2005,
	title = {Fixed-{Point} {Quantum} {Search}},
	volume = {95},
	copyright = {http://link.aps.org/licenses/aps-default-license},
	issn = {0031-9007, 1079-7114},
	url = {https://link.aps.org/doi/10.1103/PhysRevLett.95.150501},
	doi = {10.1103/PhysRevLett.95.150501},
	number = {15},
	urldate = {2024-10-15},
	journal = {Physical Review Letters},
	author = {Grover, Lov K.},
	month = oct,
	year = {2005},
	pages = {150501},
}

@article{mizel_critically_2009,
	title = {Critically {Damped} {Quantum} {Search}},
	volume = {102},
	copyright = {http://link.aps.org/licenses/aps-default-license},
	issn = {0031-9007, 1079-7114},
	url = {https://link.aps.org/doi/10.1103/PhysRevLett.102.150501},
	doi = {10.1103/PhysRevLett.102.150501},
	number = {15},
	urldate = {2024-10-22},
	journal = {Physical Review Letters},
	author = {Mizel, Ari},
	month = apr,
	year = {2009},
	pages = {150501},
}

@article{yoder_fixed-point_2014,
    author = "Yoder, Theodore J. and Low, Guang Hao and Chuang, Isaac L.",
    title = "{Fixed-Point Quantum Search with an Optimal Number of Queries}",
    doi = "10.1103/PhysRevLett.113.210501",
    journal = "Phys. Rev. Lett.",
    volume = "113",
    number = "21",
    pages = "210501",
    year = "2014"
}

@article{berry_exponential_2014,
author = {Berry, Dominic W. and Childs, Andrew M. and Cleve, Richard and Kothari, Robin and Somma, Rolando D.},
title = {Exponential improvement in precision for simulating sparse Hamiltonians},
year = {2017}, volume={5}, DOI={10.1017/fms.2017.2}, journal={Forum of Mathematics, Sigma},
}

@article{yan_fixed-point_2022,
	title = {Fixed-point oblivious quantum amplitude-amplification algorithm},
	volume = {12},
	copyright = {2022 The Author(s)},
	issn = {2045-2322},
	url = {https://www.nature.com/articles/s41598-022-15093-x},
	doi = {10.1038/s41598-022-15093-x},
	number = {1},
	urldate = {2024-10-22},
	journal = {Scientific Reports},
	author = {Yan, Bao and Wei, Shijie and Jiang, Haocong and Wang, Hong and Duan, Qianheng and Ma, Zhi and Long, Gui-Lu},
	month = aug,
	year = {2022},
	note = {Publisher: Nature Publishing Group},
	pages = {14339},
}

@misc{zecchi_improved_2025,
	doi = {10.1088/2058-9565/addeea},
url = {https://doi.org/10.1088/2058-9565/addeea},
year = {2025},
month = {jun},
publisher = {IOP Publishing},
volume = {10},
number = {3},
pages = {035039},
author = {Zecchi, Alessandro Andrea and Sanavio, Claudio and Perotto, Simona and Succi, Sauro},
title = {Improved amplitude amplification strategies for the quantum simulation of classical transport problems},
journal = {Quantum Science and Technology},
}

@inproceedings{gilyen_quantum_2019,
	address = {Phoenix AZ USA},
	title = {Quantum singular value transformation and beyond: exponential improvements for quantum matrix arithmetics},
	isbn = {978-1-4503-6705-9},
	shorttitle = {Quantum singular value transformation and beyond},
	url = {https://dl.acm.org/doi/10.1145/3313276.3316366},
	doi = {10.1145/3313276.3316366},
	urldate = {2024-10-09},
	booktitle = {Proceedings of the 51st {Annual} {ACM} {SIGACT} {Symposium} on {Theory} of {Computing}},
	publisher = {ACM},
	author = {Gilyén, András and Su, Yuan and Low, Guang Hao and Wiebe, Nathan},
	month = jun,
	year = {2019},
	pages = {193--204},
}

@article{martyn_grand_2021,
	title = {Grand {Unification} of {Quantum} {Algorithms}},
	volume = {2},
	issn = {2691-3399},
	url = {https://link.aps.org/doi/10.1103/PRXQuantum.2.040203},
	doi = {10.1103/PRXQuantum.2.040203},
	number = {4},
	urldate = {2024-10-09},
	journal = {PRX Quantum},
	author = {Martyn, John M. and Rossi, Zane M. and Tan, Andrew K. and Chuang, Isaac L.},
	month = dec,
	year = {2021},
	pages = {040203},
}

@article{quantum_for_many_body,
       author = {{Combarro}, E.~F. and {R{\'u}a}, I.~F. and {Orts}, F. and {Ortega}, G. and {Puertas}, A.~M. and {Garz{\'o}n}, E.~M.},
        title = "{Quantum algorithms to compute the neighbour list of N-body simulations}",
      journal = {Quantum Information Processing},
         year = 2024,
        month = feb,
       volume = {23},
       number = {2},
          eid = {61},
        pages = {61},
          doi = {10.1007/s11128-023-04245-1},
       adsurl = {https://ui.adsabs.harvard.edu/abs/2024QuIP...23...61C}
}

@article{Gadget,
    title = {GADGET: a code for collisionless and gasdynamical cosmological simulations},
    journal = {New Astronomy},
    volume = {6},
    number = {2},
    pages = {79-117},
    year = {2001},
    issn = {1384-1076},
    doi = {https://doi.org/10.1016/S1384-1076(01)00042-2},
    url = {https://www.sciencedirect.com/science/article/pii/S1384107601000422},
    author = {Volker Springel and Naoki Yoshida and Simon D.M. White}
}

@article{verlet_computer_1967,
	title = {Computer "{Experiments}" on {Classical} {Fluids}. {I}. {Thermodynamical} {Properties} of {Lennard}-{Jones} {Molecules}},
	volume = {159},
	copyright = {http://link.aps.org/licenses/aps-default-license},
	issn = {0031-899X},
	url = {https://link.aps.org/doi/10.1103/PhysRev.159.98},
	doi = {10.1103/PhysRev.159.98},
	number = {1},
	urldate = {2025-05-17},
	journal = {Physical Review},
	author = {Verlet, Loup},
	month = jul,
	year = {1967},
	pages = {98--103},
}

@article{Verlet_list_time,
title = {Efficient neighbor list calculation for molecular simulation of colloidal systems using graphics processing units},
journal = {Computer Physics Communications},
volume = {203},
pages = {45-52},
year = {2016},
issn = {0010-4655},
doi = {https://doi.org/10.1016/j.cpc.2016.02.003},
url = {https://www.sciencedirect.com/science/article/pii/S0010465516300182},
author = {Michael P. Howard and Joshua A. Anderson and Arash Nikoubashman and Sharon C. Glotzer and Athanassios Z. Panagiotopoulos}
}

@INPROCEEDINGS{verlet_gpu_comparison_2,
  author={Proctor, Andrew J. and Lipscomb, Tyson J. and Zou, Anqi and Anderson, Joshua A. and Cho, Samuel S.},
  booktitle={2012 ASE/IEEE International Conference on BioMedical Computing (BioMedCom)}, 
  title={{P}erformance {A}nalyses of a {P}arallel {V}erlet {N}eighbor {L}ist {A}lgorithm for {GPU}-{O}ptimized {MD} {S}imulations}, 
  year={2012},
  pages={14-19},
  doi={10.1109/BioMedCom.2012.9}
}

@article{Verlet_list1,
    title = {Improved neighbor list algorithm in molecular simulations using cell decomposition and data sorting method},
    journal = {Computer Physics Communications},
    volume = {161},
    number = {1},
    pages = {27-35},
    year = {2004},
    issn = {0010-4655},
    doi = {https://doi.org/10.1016/j.cpc.2004.04.004},
    url = {https://www.sciencedirect.com/science/article/pii/S0010465504002097},
    author = {Zhenhua Yao and Jian-Sheng Wang and Gui-Rong Liu and Min Cheng},
}

@article{SPH_OG,
    author = {Gingold, R. A. and Monaghan, J. J.},
    title = {Smoothed particle hydrodynamics: theory and application to non-spherical stars},
    journal = {Monthly Notices of the Royal Astronomical Society},
    volume = {181},
    number = {3},
    pages = {375-389},
    year = {1977},
    month = {12},
    issn = {0035-8711},
    doi = {10.1093/mnras/181.3.375},
    url = {https://doi.org/10.1093/mnras/181.3.375},
    eprint = {https://academic.oup.com/mnras/article-pdf/181/3/375/3104055/mnras181-0375.pdf},
}

@article{burnes-hut-tree,
    author = {Barnes, Josh. and Hut, Piet},
    title = {A hierarchical {O}({N} log {N}) force-calculation algorithm},
    journal = {Nature},
    year = {1986},
    month = {12},
    issn = {1476-4687},
    doi = {10.1038/324446a0},
    url = {https://doi.org/10.1038/324446a0},
}

@misc{camps_explicit_2023,
	author = {Camps, Daan and Lin, Lin and Van Beeumen, Roel and Yang, Chao},
title = {Explicit Quantum Circuits for Block Encodings of Certain Sparse Matrices},
journal = {SIAM Journal on Matrix Analysis and Applications},
volume = {45},
number = {1},
pages = {801-827},
year = {2024},
doi = {10.1137/22M1484298},
URL = {https://doi.org/10.1137/22M1484298},
}

@article{sanavio_carleman-lattice-boltzmann_2025,
	title = {Carleman-lattice-{Boltzmann} quantum circuit with matrix access oracles},
	volume = {37},
	issn = {1070-6631},
	url = {https://doi.org/10.1063/5.0254588},
	doi = {10.1063/5.0254588},
	number = {3},
	urldate = {2025-05-24},
	journal = {Physics of Fluids},
	author = {Sanavio, Claudio and Simon, William A. and Ralli, Alexis and Love, Peter and Succi, Sauro},
	month = mar,
	year = {2025},
	pages = {037123},
}

@article{castelli_decrypting_2024,
	title = {Decrypting {Allostery} in {Membrane}-{Bound} {K}-{Ras4B} {Using} {Complementary} {In} {Silico} {Approaches} {Based} on {Unbiased} {Molecular} {Dynamics} {Simulations}},
	volume = {146},
	issn = {0002-7863},
	url = {https://doi.org/10.1021/jacs.3c11396},
	doi = {10.1021/jacs.3c11396},
	number = {1},
	urldate = {2025-06-20},
	journal = {Journal of the American Chemical Society},
	author = {Castelli, Matteo and Marchetti, Filippo and Osuna, Sílvia and F. Oliveira, A. Sofia and Mulholland, Adrian J. and Serapian, Stefano A. and Colombo, Giorgio},
	month = jan,
	year = {2024},
	note = {Publisher: American Chemical Society},
	pages = {901--919},
}

@article{turau_fixed-radius_1991,
	title = {Fixed-radius near neighbors search},
	volume = {39},
	issn = {0020-0190},
	url = {https://www.sciencedirect.com/science/article/pii/002001909190180P},
	doi = {10.1016/0020-0190(91)90180-P},
	number = {4},
	urldate = {2025-06-20},
	journal = {Information Processing Letters},
	author = {Turau, Volker},
	month = aug,
	year = {1991},
	pages = {201--203},
}

@misc{chen_fast_2024,
title={Fast and exact fixed-radius neighbor search based on sorting},
  author={Chen, Xinye and G{\"u}ttel, Stefan},
  journal={PeerJ Computer Science},
  volume={10},
  pages={e1929},
  year={2024},
  publisher={PeerJ Inc.},
  doi={ https://doi.org/10.7717/peerj-cs.1929}
}

@article{Zhang_2022,
doi = {10.1209/0295-5075/ac90e6},
url = {https://dx.doi.org/10.1209/0295-5075/ac90e6},
year = {2022},
month = {sep},
publisher = {EDP Sciences, IOP Publishing and Società Italiana di Fisica},
volume = {140},
number = {1},
pages = {18002},
author = {Zhang, K. and Yu, K. and Korepin, V.},
title = {Quantum search on noisy intermediate-scale quantum devices},
journal = {Europhysics Letters}
}

@article{PhysRevResearch.7.L012017,
  title = {Noise-tolerant {G}rover's algorithm via success-probability prediction},
  author = {Leng, Jian and Yang, Fan and Wang, Xiang-Bin},
  journal = {Phys. Rev. Res.},
  volume = {7},
  issue = {1},
  pages = {L012017},
  numpages = {7},
  year = {2025},
  month = {Jan},
  publisher = {American Physical Society},
  doi = {10.1103/PhysRevResearch.7.L012017},
  url = {https://link.aps.org/doi/10.1103/PhysRevResearch.7.L012017}
}

@article{figgatt_complete_2017,
	title = {Complete 3-{Qubit} {Grover} search on a programmable quantum computer},
	volume = {8},
	issn = {2041-1723},
	url = {https://doi.org/10.1038/s41467-017-01904-7},
	doi = {10.1038/s41467-017-01904-7},
	number = {1},
	journal = {Nature Communications},
	author = {Figgatt, C. and Maslov, D. and Landsman, K. A. and Linke, N. M. and Debnath, S. and Monroe, C.},
	month = dec,
	year = {2017},
	pages = {1918},
}

@article{malvetti_quantum_2021,
	title = {Quantum {Circuits} for {Sparse} {Isometries}},
	volume = {5},
	issn = {2521-327X},
	url = {http://arxiv.org/abs/2006.00016},
	doi = {10.22331/q-2021-03-15-412},
	language = {en},
	urldate = {2025-06-05},
	journal = {Quantum},
	author = {Malvetti, Emanuel and Iten, Raban and Colbeck, Roger},
	month = mar,
	year = {2021},
	note = {arXiv:2006.00016 [quant-ph]},
	pages = {412},
}

@article{mozafari_efficient_2021,
	title = {Efficient {Boolean} {Methods} for {Preparing} {Uniform} {Quantum} {States}},
	volume = {2},
	issn = {2689-1808},
	url = {https://ieeexplore.ieee.org/document/9506863},
	doi = {10.1109/TQE.2021.3101663},
	urldate = {2025-06-07},
	journal = {IEEE Transactions on Quantum Engineering},
	author = {Mozafari, Fereshte and Riener, Heinz and Soeken, Mathias and De Micheli, Giovanni},
	year = {2021},
	pages = {1--12},
}

@article{zhang_low-depth_2021,
	title = {Low-depth quantum state preparation},
	volume = {3},
	url = {https://link.aps.org/doi/10.1103/PhysRevResearch.3.043200},
	doi = {10.1103/PhysRevResearch.3.043200},
	number = {4},
	urldate = {2025-06-06},
	journal = {Physical Review Research},
	author = {Zhang, Xiao-Ming and Yung, Man-Hong and Yuan, Xiao},
	month = dec,
	year = {2021},
	note = {Publisher: American Physical Society},
	pages = {043200},
}

@article{zhang_quantum_2022,
	title = {Quantum {State} {Preparation} with {Optimal} {Circuit} {Depth}: {Implementations} and {Applications}},
	volume = {129},
	shorttitle = {Quantum {State} {Preparation} with {Optimal} {Circuit} {Depth}},
	url = {https://link.aps.org/doi/10.1103/PhysRevLett.129.230504},
	doi = {10.1103/PhysRevLett.129.230504},
	number = {23},
	urldate = {2025-06-06},
	journal = {Physical Review Letters},
	author = {Zhang, Xiao-Ming and Li, Tongyang and Yuan, Xiao},
	month = nov,
	year = {2022},
	note = {Publisher: American Physical Society},
	pages = {230504},
}

@article{zhang_circuit_2024,
	title = {Circuit complexity of quantum access models for encoding classical data},
	volume = {10},
	copyright = {2024 The Author(s)},
	issn = {2056-6387},
	url = {https://www.nature.com/articles/s41534-024-00835-8},
	doi = {10.1038/s41534-024-00835-8},
	language = {en},
	number = {1},
	urldate = {2025-06-06},
	journal = {npj Quantum Information},
	author = {Zhang, Xiao-Ming and Yuan, Xiao},
	month = apr,
	year = {2024},
	note = {Publisher: Nature Publishing Group},
	pages = {1--12},
}

@misc{cuccaro_new_2004,
	title = {A new quantum ripple-carry addition circuit},
	url = {https://arxiv.org/abs/quant-ph/0410184v1},
	language = {en},
	urldate = {2025-06-10},
	journal = {arXiv.org},
	author = {Cuccaro, Steven A. and Draper, Thomas G. and Kutin, Samuel A. and Moulton, David Petrie},
	month = oct,
	year = {2004},
}

@article{logarithmic_depth_quantum_adder,
author = {Draper, Thomas G. and Kutin, Samuel A. and Rains, Eric M. and Svore, Krysta M.},
title = {A logarithmic-depth quantum carry-lookahead adder},
year = {2006},
issue_date = {July 2006},
publisher = {Rinton Press, Incorporated},
address = {Paramus, NJ},
volume = {6},
number = {4},
issn = {1533-7146},
journal = {Quantum Info. Comput.},
month = jul,
pages = {351–369},
numpages = {19}
}

@misc{egretta_thula,
  title = {Quantum Incrementer},
  howpublished = {https://egrettathula.wordpress.com/2024/07/28/quantum-incrementer/},
  author = {Egretta Thula},
  note = {July 28 2024}
}

@article{vale_circuit_2024,
	title = {Circuit {Decomposition} of {Multicontrolled} {Special} {Unitary} {Single}-{Qubit} {Gates}},
	volume = {43},
	issn = {1937-4151},
	url = {https://ieeexplore.ieee.org/document/10293178},
	doi = {10.1109/TCAD.2023.3327102},
	number = {3},
	urldate = {2025-01-30},
	journal = {IEEE Transactions on Computer-Aided Design of Integrated Circuits and Systems},
	author = {Vale, Rafaella and Azevedo, Thiago Melo D. and Araújo, Ismael C. S. and Araujo, Israel F. and da Silva, Adenilton J.},
	month = mar,
	year = {2024},
	note = {Conference Name: IEEE Transactions on Computer-Aided Design of Integrated Circuits and Systems},
	pages = {802--811},
}

@misc{zindorf_efficient_2025,
	title = {Efficient {Implementation} of {Multi}-{Controlled} {Quantum} {Gates}},
	url = {http://arxiv.org/abs/2404.02279},
	doi = {10.48550/arXiv.2404.02279},
	urldate = {2025-06-10},
	publisher = {arXiv},
	author = {Zindorf, Ben and Bose, Sougato},
	month = mar,
	year = {2025},
	note = {arXiv:2404.02279 [quant-ph]},
}

@article{diagonal_gate_Shende_2006,
   title={Synthesis of quantum-logic circuits},
   volume={25},
   ISSN={1937-4151},
   url={http://dx.doi.org/10.1109/TCAD.2005.855930},
   DOI={10.1109/tcad.2005.855930},
   number={6},
   journal={IEEE Transactions on Computer-Aided Design of Integrated Circuits and Systems},
   publisher={Institute of Electrical and Electronics Engineers (IEEE)},
   author={Shende, V.V. and Bullock, S.S. and Markov, I.L.},
   year={2006},
   month=jun, pages={1000–1010} }

@Article{electronics13234830,
AUTHOR = {Lee, Sihyung and Nam, Seung Yeob},
TITLE = {Finding All Solutions with Grover’s Algorithm by Integrating Estimation and Discovery},
JOURNAL = {Electronics},
VOLUME = {13},
YEAR = {2024},
NUMBER = {23},
ARTICLE-NUMBER = {4830},
URL = {https://www.mdpi.com/2079-9292/13/23/4830},
ISSN = {2079-9292},
DOI = {10.3390/electronics13234830}
}

@inproceedings{brassard1998quantum,
  title={Quantum counting},
  author={Brassard, Gilles and H{\o}yer, Peter and Tapp, Alain},
  booktitle={Automata, Languages and Programming: 25th International Colloquium, ICALP'98 Aalborg, Denmark, July 13--17, 1998 Proceedings 25},
  pages={820--831},
  year={1998},
  organization={Springer}
}

@misc{superpc1,
    title={supercomputer fugaku - fujitsu},
    author = {},
    url = {https://www.fujitsu.com/global/about/innovation/fugaku/specifications/},
}

@misc{superpc2,
    title={Summit Architecture Overview},
    author = {Tom Papatheodore},
    url = {https://www.olcf.ornl.gov/wp-content/uploads/2019/05/Summit_System_Overview_20190520.pdf?utm_source=chatgpt.com
},
}

@article{SPH_neighbors,
    author = {Read, J. I. and Hayfield, T.},
    title = {SPHS: smoothed particle hydrodynamics with a higher order dissipation switch},
    journal = {Monthly Notices of the Royal Astronomical Society},
    volume = {422},
    number = {4},
    pages = {3037-3055},
    year = {2012},
    month = {05},
    issn = {0035-8711},
    doi = {10.1111/j.1365-2966.2012.20819.x},
    url = {https://doi.org/10.1111/j.1365-2966.2012.20819.x},
    eprint = {https://academic.oup.com/mnras/article-pdf/422/4/3037/18598311/mnras0422-3037.pdf},
}

@book{Frenkel2002,
  author    = {Daan Frenkel and Berend Smit},
  title     = {Understanding Molecular Simulation: From Algorithms to Applications},
  doi = {https://doi.org/10.1016/B978-0-12-267351-1.X5000-7},
  edition   = {2nd},
  publisher = {Academic Press},
  year      = {2002},
  chapter   = {3},
}

@article{gromacs,
title = {GROMACS: High performance molecular simulations through multi-level parallelism from laptops to supercomputers},
journal = {SoftwareX},
volume = {1-2},
pages = {19-25},
year = {2015},
issn = {2352-7110},
doi = {https://doi.org/10.1016/j.softx.2015.06.001},
url = {https://www.sciencedirect.com/science/article/pii/S2352711015000059},
author = {Mark James Abraham and Teemu Murtola and Roland Schulz and Szilárd Páll and Jeremy C. Smith and Berk Hess and Erik Lindahl}
}

@article{sph_today_time,
    doi = {10.1371/journal.pone.0020685},
    author = {Crespo, Alejandro C. AND Dominguez, Jose M. AND Barreiro, Anxo AND Gómez-Gesteira, Moncho AND Rogers, Benedict D.},
    journal = {PLOS ONE},
    publisher = {Public Library of Science},
    title = {GPUs, a New Tool of Acceleration in CFD: Efficiency and Reliability on Smoothed Particle Hydrodynamics Methods},
    year = {2011},
    month = {06},
    volume = {6},
    url = {https://doi.org/10.1371/journal.pone.0020685},
    pages = {1-13},
    number = {6},
}

@INPROCEEDINGS{tree_traversal_time,
  author={Cassell, Thomas Lane and Deakin, Tom and Alpay, Aksel and Heuveline, Vincent and Gadeschi, Gonzalo Brito},
  booktitle={SC24-W: Workshops of the International Conference for High Performance Computing, Networking, Storage and Analysis}, 
  title={Efficient Tree-based Parallel Algorithms for N-Body Simulations Using C++ Standard Parallelism}, 
  year={2024},
  volume={},
  number={},
  pages={708-717},
  doi={10.1109/SCW63240.2024.00099}}

@article{membrane_dynamics1,
    author = {Bond, P. J. and Sansom, M. S.},
    title = {Insertion and assembly of membrane proteins via simulation},
    journal = {Journal of the American Chemical Society} ,
    year = {2006},
    doi = {10.1021/ja0569104}
}

@article{drug_binding1,
    author = {Wang, Lingle and Wu, Yujie and Deng, Yuqing and Kim, Byungchan and Pierce, Levi and Krilov, Goran and Lupyan, Dmitry and Robinson, Shaughnessy and Dahlgren, Markus K. and Greenwood, Jeremy and Romero, Donna L. and Masse, Craig and Knight, Jennifer L. and Steinbrecher, Thomas and Beuming, Thijs and Damm, Wolfgang and Harder, Ed and Sherman, Woody and Brewer, Mark and Wester, Ron and Murcko, Mark and Frye, Leah and Farid, Ramy and Lin, Teng and Mobley, David L. and Jorgensen, William L. and Berne, Bruce J. and Friesner, Richard A. and Abel, Robert},
    title = {Accurate and Reliable Prediction of Relative Ligand Binding Potency in Prospective Drug Discovery by Way of a Modern Free-Energy Calculation Protocol and Force Field},
    journal = {Journal of the American Chemical Society},
    volume = {137},
    number = {7},
    pages = {2695-2703},
    year = {2015},
    doi = {10.1021/ja512751q},
}

@article{drug_binding2,
    author = {Shan, Yibing and Kim, Eric T. and Eastwood, Michael P. and Dror, Ron O. and Seeliger, Markus A. and Shaw, David E.},
    title = {How Does a Drug Molecule Find Its Target Binding Site?},
    journal = {Journal of the American Chemical Society},
    volume = {133},
    number = {24},
    pages = {9181-9183},
    year = {2011},
    doi = {10.1021/ja202726y},
    URL = {https://doi.org/10.1021/ja202726y}
}

@article{polymer_dynamics,
  author       = {Hengheng Zhao and Pengwei Duan and Zhenyuan Li and Qionghai Chen and Tongkui Yue and Liqun Zhang and Venkat Ganesan and Jun Liu},
  title        = {Unveiling the Multiscale Dynamics of Polymer Vitrimers Via Molecular Dynamics Simulations},
  journal      = {Macromolecules},
  year         = {2023},
  volume       = {56},
  number       = {23},
  pages        = {9336--9349},
  doi          = {10.1021/acs.macromol.3c01893},
}

@article{membrane_dynamic_2,
  author       = {Liu, Chunhong and Xue, Lingfeng and Song, Chen},
  title        = {Calcium binding and permeation in TRPV channels: Insights from molecular dynamics simulations},
  journal      = {Journal of General Physiology},
  year         = {2023},
  volume       = {155},
  number       = {12},
  pages        = {e202213261},
  doi          = {10.1085/jgp.202213261},
  note         = {Epub 2023 Sep 20},
}

@article{alphafold3,
  author       = {Abramson, Josh and Adler, Jonas and Dunger, Jack and Evans, Richard and Green, Tim and Pritzel, Alexander and Ronneberger, Olaf and Willmore, Lindsay and Ballard, Andrew J. and Bambrick, Joshua and Bodenstein, Sebastian W. and Evans, David A. and Hung, Chia-Chun and O’Neill, Michael and Reiman, David and Tunyasuvunakool, Kathryn and Wu, Zachary and Žemgulytė, Akvilė and Arvaniti, Eirini and Beattie, Charles and Bertolli, Ottavia and Bridgland, Alex and Cherepanov, Alexey and Congreve, Miles and Cowen-Rivers, Alexander I. and Cowie, Andrew and Figurnov, Michael and Fuchs, Fabian B. and Gladman, Hannah and Jain, Rishub and Khan, Yousuf A. and Low, Caroline M. R. and Perlin, Kuba and Potapenko, Anna and Savy, Pascal and Singh, Sukhdeep and Stecula, Adrian and Thillaisundaram, Ashok and Tong, Catherine and Yakneen, Sergei and Zhong, Ellen D. and Zielinski, Michal and Žídek, Augustin and Bapst, Victor and Kohli, Pushmeet and Jaderberg, Max and Hassabis, Demis and Jumper, John M.},
  title        = {Accurate structure prediction of biomolecular interactions with AlphaFold 3},
  journal      = {Nature},
  year         = {2024},
  volume       = {630},
  number       = {8016},
  pages        = {493--500},
  doi          = {10.1038/s41586-024-07487-w},
}

@article{homology1,
  author       = {Sch{\"u}tze, Konstantin and Heinzinger, Michael and Steinegger, Martin and Rost, Burkhard},
  title        = {Nearest neighbor search on embeddings rapidly identifies distant protein relations},
  journal      = {Frontiers in Bioinformatics},
  year         = {2022},
  volume       = {2},
  pages        = {1033775},
  doi          = {10.3389/fbinf.2022.1033775},
}

@article{homology2,
  author       = {van Kempen, Michel and Kim, Stephanie S. and Tumescheit, Charlotte and Mirdita, Milot and Lee, Jeongjae and Gilchrist, Cameron L. M. and S\"oding, Johannes and Steinegger, Martin},
  title        = {Fast and accurate protein structure search with Foldseek},
  journal      = {Nature Biotechnology},
  year         = {2024},
  volume       = {42},
  number       = {2},
  pages        = {243--246},
  doi          = {10.1038/s41587-023-01773-0},
}

@article{LindorffLarsen2011,
  author       = {Lindorff-Larsen, Kresten and Piana, Stefano and Dror, Ron O. and Shaw, David E.},
  title        = {How Fast-Folding Proteins Fold},
  journal      = {Science},
  year         = {2011},
  volume       = {334},
  number       = {6055},
  pages        = {517--520},
  doi          = {10.1126/science.1208351}
}

@article{Nonequilibrium1pf,
  author       = {De Los Rios, Paolo and Barducci, Alessandro},
  title        = {Non-equilibrium protein folding and activation by ATP-driven chaperones},
  journal      = {Nature Reviews Molecular Cell Biology},
  year         = {2014},
  volume       = {15},
  number       = {10},
  pages        = {611--622},
  doi          = {10.1038/nrm3868}
}

@article{non_eq_pf_2,
  author       = {Goloubinoff, Pierre and Sulpizi, Marialore and Cappellaro, Andrea and Dutheil, Julien and Nguyen, Hoang Anh and De Los Rios, Paolo},
  title        = {Chaperones convert the energy from ATP into the nonequilibrium stabilization of native proteins},
  journal      = {Nature Chemical Biology},
  year         = {2018},
  volume       = {14},
  number       = {4},
  pages        = {388--395},
  doi          = {10.1038/s41589-018-0013-8}
}

@article{non_eq_pf_3,
  author       = {Assenza, Salvatore and Sassi, Andrea S. and Kellner, Robert and De Los Rios, Paolo and Barducci, Alessandro},
  title        = {Efficient conversion of chemical energy into mechanical work by Hsp70 chaperones},
  journal      = {Physical Review X},
  year         = {2019},
  volume       = {9},
  number       = {4},
  pages        = {041033},
  doi          = {10.1103/PhysRevX.9.041033}
}

@article{protein_folding_and_neuro_diseases,
  author       = {Chiti, Fabrizio and Dobson, Christopher M.},
  title        = {Protein Misfolding, Amyloid Formation, and Human Disease: A Summary of Progress Over the Last Decade},
  journal      = {Annual Review of Biochemistry},
  year         = {2017},
  volume       = {86},
  pages        = {27--68},
  doi          = {10.1146/annurev-biochem-061516-045115}
}

@article{non-eq-for-neuro1,
  author       = {Sontag, Emily M. and Samant, Rahul S. and Frydman, Judith},
  title        = {Mechanisms and Functions of Spatial Protein Quality Control},
  journal      = {Annual Review of Biochemistry},
  year         = {2017},
  volume       = {86},
  pages        = {97--122},
  doi          = {10.1146/annurev-biochem-060815-014616}
}

@article{non-eq-for-neuro2,
  author       = {Klaips, Courtney L. and Jayaraj, Gopal G. and Hartl, F. Ulrich},
  title        = {Pathways of cellular proteostasis in aging and disease},
  journal      = {Journal of Cell Biology},
  year         = {2018},
  volume       = {217},
  number       = {1},
  pages        = {51--63},
  doi          = {10.1083/jcb.201709072}
}

@article{Succi_2023_overview,
doi = {10.1209/0295-5075/acfdc7},
url = {https://dx.doi.org/10.1209/0295-5075/acfdc7},
year = {2023},
month = {oct},
publisher = {EDP Sciences, IOP Publishing and Società Italiana di Fisica},
volume = {144},
number = {1},
pages = {10001},
author = {Succi, Sauro and Itani, W. and Sreenivasan, K. and Steijl, R.},
title = {Quantum computing for fluids: Where do we stand?},
journal = {Europhysics Letters}
}

@article{DiIlio_fluiddynamics_2018,
title = {Fluid flow around NACA 0012 airfoil at low-Reynolds numbers with hybrid lattice Boltzmann method},
journal = {Computers \& Fluids},
volume = {166},
pages = {200-208},
year = {2018},
issn = {0045-7930},
doi = {https://doi.org/10.1016/j.compfluid.2018.02.014},
url = {https://www.sciencedirect.com/science/article/pii/S0045793018300677},
author = {G. {Di Ilio} and D. Chiappini and S. Ubertini and G. Bella and S. Succi},
}

@article{cappelli_vlasov_2024,
  title = {From Vlasov-Poisson to Schr\"odinger-Poisson: Dark matter simulation with a quantum variational time evolution algorithm},
  author = {Cappelli, Luca and Tacchino, Francesco and Murante, Giuseppe and Borgani, Stefano and Tavernelli, Ivano},
  journal = {Phys. Rev. Res.},
  volume = {6},
  issue = {1},
  pages = {013282},
  numpages = {16},
  year = {2024},
  month = {Mar},
  publisher = {American Physical Society},
  doi = {10.1103/PhysRevResearch.6.013282},
  url = {https://link.aps.org/doi/10.1103/PhysRevResearch.6.013282}
}

@article{cao_quantum_chemistry_2019,
author = {Cao, Yudong and Romero, Jonathan and Olson, Jonathan P. and Degroote, Matthias and Johnson, Peter D. and Kieferová, Mária and Kivlichan, Ian D. and Menke, Tim and Peropadre, Borja and Sawaya, Nicolas P. D. and Sim, Sukin and Veis, Libor and Aspuru-Guzik, Alán},
title = {Quantum Chemistry in the Age of Quantum Computing},
journal = {Chemical Reviews},
volume = {119},
number = {19},
pages = {10856-10915},
year = {2019},
doi = {10.1021/acs.chemrev.8b00803},
}

@article{alexeev_quantum-centric_2024,
	title = {Quantum-centric supercomputing for materials science: A perspective on challenges and future directions},
	volume = {160},
	issn = {0167-739X},
	shorttitle = {Quantum-centric supercomputing for materials science},
	url = {https://www.sciencedirect.com/science/article/pii/S0167739X24002012},
	doi = {10.1016/j.future.2024.04.060},
	urldate = {2025-01-22},
	journal = {Future Generation Computer Systems},
	author = {Alexeev, Yuri and Amsler, Maximilian and Barroca, Marco Antonio and Bassini, Sanzio and Battelle, Torey and Camps, Daan and Casanova, David and Choi, Young Jay and Chong, Frederic T. and Chung, Charles and Codella, Christopher and Córcoles, Antonio D. and Cruise, James and Di Meglio, Alberto and Duran, Ivan and Eckl, Thomas and Economou, Sophia and Eidenbenz, Stephan and Elmegreen, Bruce and Fare, Clyde and Faro, Ismael and Fernández, Cristina Sanz and Ferreira, Rodrigo Neumann Barros and Fuji, Keisuke and Fuller, Bryce and Gagliardi, Laura and Galli, Giulia and Glick, Jennifer R. and Gobbi, Isacco and Gokhale, Pranav and de la Puente Gonzalez, Salvador and Greiner, Johannes and Gropp, Bill and Grossi, Michele and Gull, Emanuel and Healy, Burns and Hermes, Matthew R. and Huang, Benchen and Humble, Travis S. and Ito, Nobuyasu and Izmaylov, Artur F. and Javadi-Abhari, Ali and Jennewein, Douglas and Jha, Shantenu and Jiang, Liang and Jones, Barbara and de Jong, Wibe Albert and Jurcevic, Petar and Kirby, William and Kister, Stefan and Kitagawa, Masahiro and Klassen, Joel and Klymko, Katherine and Koh, Kwangwon and Kondo, Masaaki and K\"urk\c{c}\"u\u{g}lu, Do\u{g}a Murat and Kurowski, Krzysztof and Laino, Teodoro and Landfield, Ryan and Leininger, Matt and Leyton-Ortega, Vicente and Li, Ang and Lin, Meifeng and Liu, Junyu and Lorente, Nicolas and Luckow, Andre and Martiel, Simon and Martin-Fernandez, Francisco and Martonosi, Margaret and Marvinney, Claire and Medina, Arcesio Castaneda and Merten, Dirk and Mezzacapo, Antonio and Michielsen, Kristel and Mitra, Abhishek and Mittal, Tushar and Moon, Kyungsun and Moore, Joel and Mostame, Sarah and Motta, Mario and Na, Young-Hye and Nam, Yunseong and Narang, Prineha and Ohnishi, Yu-ya and Ottaviani, Daniele and Otten, Matthew and Pakin, Scott and Pascuzzi, Vincent R. and Pednault, Edwin and Piontek, Tomasz and Pitera, Jed and Rall, Patrick and Ravi, Gokul Subramanian and Robertson, Niall and Rossi, Matteo A. C. and Rydlichowski, Piotr and Ryu, Hoon and Samsonidze, Georgy and Sato, Mitsuhisa and Saurabh, Nishant and Sharma, Vidushi and Sharma, Kunal and Shin, Soyoung and Slessman, George and Steiner, Mathias and Sitdikov, Iskandar and Suh, In-Saeng and Switzer, Eric D. and Tang, Wei and Thompson, Joel and Todo, Synge and Tran, Minh C. and Trenev, Dimitar and Trott, Christian and Tseng, Huan-Hsin and Tubman, Norm M. and Tureci, Esin and Valinas, David García and Vallecorsa, Sofia and Wever, Christopher and Wojciechowski, Konrad and Wu, Xiaodi and Yoo, Shinjae and Yoshioka, Nobuyuki and Yu, Victor Wen-zhe and Yunoki, Seiji and Zhuk, Sergiy and Zubarev, Dmitry},
	month = nov,
	year = {2024},
	pages = {666--710},
}

@techreport{eurex,
    author = {Daniele Cesarini and Fabio Pitari and Filippo Barbari and Federico Ficarelli and Piero Lanucara and Gabriele Cavallaro and Emanuele Casarotti and Igor Piljic´ and Andrew Beggs and Olivier Marsden and Ioan Hadade and Luca Tornatore and Vasilis Flouris and Fotis Nikolaidis and Angelos Bilas and Erwan Raffin and Antoine Morvan and Ondrej Meca and Riha Lubomir and Theophile Lohier and Valerie Brenner and Luigi Genovese and Maxime Delorme},
    title = {EUPEX deliverable D3.2 Applications optimised for SVE and HBM},
    institution = {EuropeanUnion},
    year = {2023} 
}

@article{Habib2017HACC,
  author    = {Salman Habib and Vitali Morozov and Nicholas Frontiere and Hal Finkel and Adrian Pope and Katrin Heitmann and Kalyan Kumaran and Venkatram Vishwanath and Tom Peterka and Joe Insley and David Daniel and Patricia Fasel and Zarija Luki{\'c}},
  title     = {{HACC}: Extreme Scaling and Performance Across Diverse Architectures},
  journal   = {Commun.\ ACM},
  year      = {2017},
  volume    = {60},
  number    = {1},
  pages     = {97-104},
  doi       = {10.1145/3015569},  
  url       = {https://cacm.acm.org/research/hacc/}
}

@article{qram,
  title={QRAM: A Survey and Critique},
  author={Jaques, Samuel and Rattew, Arthur G},
  journal={Quantum},
  volume={9},
  pages={1922},
  year={2025},
  publisher={Verein zur F{\"o}rderung des Open Access Publizierens in den Quantenwissenschaften}, 
doi = {	https://doi.org/10.22331/q-2025-12-02-1922}
}

@book{feder2022,
  title={Physics of flow in porous media},
  author={Feder, Jens and Flekk{\o}y, Eirik Grude and Hansen, Alex},
  year={2022},
  publisher={Cambridge University Press}
}

@article{sahimi1993,
  title={Flow phenomena in rocks: from continuum models to fractals, percolation, cellular automata, and simulated annealing},
  author={Sahimi, Muhammad},
  journal={Reviews of modern physics},
  volume={65},
  number={4},
  pages={1393},
  year={1993},
  publisher={APS},
  doi= {https://doi.org/10.1103/RevModPhys.65.1393}
}

@article{cali1992,
  title={Diffusion and hydrodynamic dispersion with the lattice Boltzmann method},
  author={Cali, A and Succi, S and Cancelliere, A and Benzi, R and Gramignani, M},
  journal={Physical Review A},
  volume={45},
  number={8},
  pages={5771},
  year={1992},
  doi = {https://doi.org/10.1103/PhysRevA.45.5771},
  publisher={APS}
}

@article{KLINKENBERG2023,
title = {H2M: Exploiting Heterogeneous Shared Memory Architectures},
journal = {Future Generation Computer Systems},
volume = {148},
pages = {39-55},
year = {2023},
issn = {0167-739X},
doi = {https://doi.org/10.1016/j.future.2023.05.019},
url = {https://www.sciencedirect.com/science/article/pii/S0167739X23002029},
author = {Jannis Klinkenberg and Anara Kozhokanova and Christian Terboven and Clément Foyer and Brice Goglin and Emmanuel Jeannot}
}

@article{DRAVAI2025,
title = {Performance and efficiency: A multi-generational benchmark of modern processors on bandwidth-bound HPC applications},
journal = {Future Generation Computer Systems},
volume = {169},
pages = {107793},
year = {2025},
issn = {0167-739X},
doi = {https://doi.org/10.1016/j.future.2025.107793},
url = {https://www.sciencedirect.com/science/article/pii/S0167739X25000883},
author = {Balázs Drávai and István Z. Reguly}
}

@article{springel2005cosmological,
  title={The cosmological simulation code GADGET-2},
  author={Springel, Volker},
  journal={Monthly notices of the royal astronomical society},
  volume={364},
  number={4},
  pages={1105--1134},
  year={2005},
  publisher={The Royal Astronomical Society}
}

@article{zhu2015numerical,
  title={Numerical convergence in smoothed particle hydrodynamics},
  author={Zhu, Qirong and Hernquist, Lars and Li, Yuexing},
  journal={The Astrophysical Journal},
  volume={800},
  number={1},
  pages={6},
  year={2015},
  publisher={IOP Publishing},
  doi = {10.1088/0004-637X/800/1/6}
}

@article{alves2022cache,
  title={Cache-oblivious hilbert curve-based blocking scheme for matrix transposition},
  author={Alves, Jo{\~a}o Nuno Ferreira and Russo, Lu{\'\i}s Manuel Silveira and Francisco, Alexandre},
  journal={ACM Transactions on Mathematical Software},
  volume={48},
  number={4},
  pages={1--28},
  year={2022},
  publisher={ACM New York, NY},
doi = {https://doi.org/10.1145/3555353},
}

\end{document}